\documentclass[acmsmall,nonacm]{acmart}

\usepackage{tikz}
\usetikzlibrary{arrows.meta, positioning, shapes.geometric, backgrounds, fit}
\usepackage{xspace}
\usepackage[nointegrals]{wasysym}
\usepackage{longtable}
\usepackage{array}
\usepackage{colortbl}
\definecolor{gridgray}{gray}{0.7}
\newcolumntype{g}{!{\color{gridgray}\vrule}}
\newcolumntype{b}{!{\color{black}\vrule}}
\newcolumntype{L}[1]{>{\raggedright\arraybackslash}p{#1}}
\newcommand{\abci}{\texttt{ABCI}\xspace}
\newcommand{\checktx}{\texttt{CheckTx}\xspace}

\newcommand{\statevar}[1]{\texttt{#1}}
\newcommand{\term}[1]{\emph{#1}}

\title{SoK: Design Choices in Tendermint BFT Consensus}

\author{Sanaz Taheri-Boshrooyeh}
\affiliation{%
  \institution{MathCrypt Research Inc.}
  \city{Vancouver}
  \country{Canada}
}
\email{sanaz@mathcrypt.com}

\author{Yahya Hassanzadeh-Nazarabadi}
\affiliation{%
  \institution{MathCrypt Research Inc.}
  \city{Vancouver}
  \country{Canada}
}
\email{yahya@mathcrypt.com}

\begin{document}

\begin{abstract}
Tendermint is a consensus protocol that tolerates Byzantine faults, implemented first by the
Tendermint Core engine and now by CometBFT. Realizing the protocol's separation of agreement from
execution, the Application Blockchain Interface (ABCI) of these engines keeps consensus apart from the replicated application, so one engine can serve
many independent proof-of-stake chains. A sustained body of research proposes changes that span the
protocol's mechanisms, from vote dissemination to block proposal and reward distribution, yet to our knowledge no survey studies the protocol and these works as one family,
and broader surveys do not take Tendermint as their subject. We describe the baseline, the protocol as
consolidated from its original descriptions, by its design choices. Each is a decision, such as how
votes spread or how proposers rotate, through which the baseline provides safety and liveness or another
property under its fault and timing model. We classify each work by the design choice
it renegotiates, that is, modifies or replaces, by the properties it strengthens or weakens, and by whether its result needs a stronger model.
Taking a broader perspective on that classification, we group the surveyed works' solutions
into a few threads of concern in the baseline, namely its performance, the
accountability of its validators, and the protection of exploitable information. For each thread, we identify
the recurring strategies that their approaches follow. Together, these threads and strategies describe
the current state of research on the Tendermint family. Where the strategies leave gaps, we pose
future directions, each as an open question under an explicit fault and timing
model.

\end{abstract}

\keywords{Tendermint, CometBFT, Byzantine fault tolerance, blockchain consensus, proof of stake, ABCI, Cosmos}

\maketitle

\section{Introduction}
\label{sec:intro}

Tendermint is a Byzantine fault tolerant (BFT) consensus protocol, and its current
implementation is CometBFT~\cite{cometbft}, the successor to the original Tendermint Core
engine~\cite{tendermintcore}. Both engines perform state-machine replication, keeping
a set of validators in agreement on one ordered log of transactions. The Tendermint
protocol has been widely deployed through its engines~\cite{cason2021design}, including in Osmosis~\cite{osmosis}, Celestia~\cite{celestia}, the dYdX
Chain~\cite{dydxchain}, and Akash~\cite{akash}, which run CometBFT or forks of it.
Since 2019, the protocol has also secured the Cosmos Hub~\cite{cosmoshub}, one of the largest
deployments of a BFT state-machine replication platform as of 2022~\cite{buchman2022revisiting}, running
first in Tendermint Core and then in CometBFT.
Each engine separates consensus from the
application, the state machine that executes the ordered transactions. The
Application Blockchain Interface (\abci) is the boundary between the two. An
application communicates with the consensus core over \abci{}, so the same engine
can replicate different applications~\cite{buchman2022revisiting}. Because \abci{}
can also run over a socket between separate processes, an application can be
written in any programming language~\cite{cometbft,buchman2016tendermint}. Building
on \abci{}, the Cosmos SDK~\cite{cosmossdk} packages the application layer and eases
application development on the engine. Through \abci{} and the Cosmos SDK, the
Tendermint protocol underpins many independent, application-specific proof-of-stake
chains.

Alongside this deployment, a line of research has proposed changes to Tendermint, each
intended to improve the protocol in some respect.
To the best of our knowledge, no survey is dedicated to Tendermint and the works built on it.
The nearest surveys cover blockchain systems or BFT consensus in general and mention Tendermint only as one protocol
among many, not as their subject.
One such survey~\cite{xu2024exploring} analyzes blockchain systems through a modular framework of network, consensus, and distributed-ledger layers. It assigns Tendermint to the consensus layer, as one BFT protocol beside PBFT and HotStuff.
Likewise, a review of BFT consensus algorithms~\cite{zhang2024reaching} surveys designs such as PBFT, HotStuff, and protocols based on directed acyclic graphs, and cites Tendermint only in passing.
As a result, neither survey examines the changes that the works built on Tendermint propose to it.
The wide adoption of Tendermint through its engines makes these changes worth a survey of their own.
For each mechanism of the protocol, such a survey can show what the literature has proposed, what each proposal gains, which other properties it affects, and which areas remain unexplored.

In this paper, we study the existing efforts to improve the Tendermint protocol. These works
follow the protocol's lineage, since some extend the protocol or one of its implementations
and others carry a general result from BFT or proof-of-stake consensus over to Tendermint.
We take as the \term{baseline} the Tendermint
algorithm as presented in~\cite{kwon2014tendermint,buchman2016tendermint,buchman2018latest,amoussou2019dissecting}. Following earlier design-space
and tradeoff analyses of BFT protocols~\cite{amiri2024bedrock,buchman2022revisiting}, we
describe the baseline by its \term{design choices}. These are the decisions through which
Tendermint delivers its safety and liveness guarantees and its other properties under its
fault and timing model. The round-robin proposer is one such choice. It passes the role of
proposer among the validators in a fixed order that every validator computes for itself,
giving each validator turns in proportion to its voting power. The rotation thereby provides
a choice of proposer that requires no election and no exchange of messages, but it also
makes each proposer known well before its turn. We then classify each surveyed work by the
design choice it renegotiates and by the properties it strengthens or weakens as a
result.\footnote{We focus on works that renegotiate a design choice of the baseline and
analyze or evaluate the result. Works that only analyze the baseline are discussed in
Section~\ref{sec:background} instead, and general BFT or proof-of-stake works that mention
Tendermint only in passing lie outside our scope.} The construction of~\cite{bui2023enhanced}, for
example, renegotiates the round-robin proposer by seeding the selection with a verifiable
random function, whose random output carries a proof that any validator can verify.
Proposers then cannot be predicted beyond the next height, so the change strengthens
proposer availability, the ability of the chain to keep making progress when an attacker
targets its proposers.

The first contribution of this survey is the classification of the surveyed works, given in
Section~\ref{sec:matrix}. Alongside the design choice and the properties that it assigns
to each work, the classification determines the model that the work's result requires, namely the baseline's
own fault and timing model or a stronger one that the work assumes. The second contribution
is a set of future directions drawn from the first. Each names a property that the
baseline does not provide and asks whether a renegotiation can provide it, and under which
fault and timing model.

The remainder of this paper is structured as follows. Section~\ref{sec:background} describes
the baseline protocol and the guarantees and properties it provides.
Section~\ref{sec:design-dna} develops the classification.
We first identify the design choices of the baseline and what each provides
(Section~\ref{subsec:dna-choices}). We then classify the surveyed works by the design choices they
renegotiate and the properties they affect, and we determine the model that each result requires
(Section~\ref{sec:matrix}). Finally, we synthesize these works by their aims, namely improving
performance, extending accountability, and protecting exploitable information
(Section~\ref{subsec:map-shows}).
Section~\ref{sec:conclusion} concludes the paper and states future directions.

\section{Background: the Tendermint protocol}
\label{sec:background}

Tendermint was first proposed in 2014~\cite{kwon2014tendermint}, specified and
evaluated in 2016~\cite{buchman2016tendermint}, and given its canonical formal
description in 2018~\cite{buchman2018latest}. A formal correctness analysis followed in
2019~\cite{amoussou2019dissecting}. This section consolidates these
descriptions into one presentation of Tendermint that the survey uses as its
\textit{baseline}, which is the protocol itself rather than any implementation of it. The section
first describes the protocol's mechanisms (Sections~\ref{subsec:core-mechanics}
to~\ref{subsec:mempool}), and Figure~\ref{fig:mechanics} illustrates one round of the protocol. The rest of the section builds on these mechanisms to explain
the guarantees the protocol provides and its measured, conditional, and economic properties.

Tendermint is a proof-of-stake protocol in which a set of validators together replicate one
growing chain. Each validator has \emph{voting power}, the weight its votes carry, proportional
to the stake it has bonded~\cite{kwon2014tendermint,buchman2016tendermint}.
Each validator is identified by a public key listed in the chain's state and signs
every message it sends, and signatures are assumed to be
unforgeable~\cite{buchman2016tendermint,buchman2018latest,amoussou2019dissecting}.
The protocol is analyzed under a model with two parts~\cite{buchman2018latest}.
The first is the fault bound, which states that a Byzantine adversary
controls strictly less than one-third of the total voting power.
The bound applies at every point of an execution, so it holds across changes to the
validator set~\cite{amoussou2019dissecting}.
The second is partial synchrony, in which the network may be arbitrarily slow for a while but
eventually stabilizes~\cite{dwork1988consensus}. That is, after an unknown instant, the \term{Global
Stabilization Time} (GST), every
message that one correct validator sends to another arrives within a bounded time.
Beyond these two parts, Tendermint requires that whatever one correct validator has received, every correct
validator also receives within a bounded time after GST. This requirement holds even for a message that a
Byzantine validator started, once any correct validator holds it. This requirement is called the
\emph{gossip communication property}~\cite{buchman2018latest}.
Under the fault bound, partial synchrony, and the gossip communication property, the protocol
provides the safety and liveness guarantees of Section~\ref{subsec:guarantees}.

Both guarantees rest on the size of a \term{quorum}, the share of voting power that a voting
step needs. Derived from the fault bound, it is more than two-thirds of the total voting power. A quorum
of that size is large enough that any two quorums overlap in more than one-third of the voting
power and so share a correct validator, a property called
\term{quorum intersection}~\cite{buchman2018latest}. Such a quorum is also small enough that the correct
validators alone can form one.

\subsection{Core mechanics}
\label{subsec:core-mechanics}
\begin{figure*}[t]
\centering
\tikzset{
  tmd/.style={font=\footnotesize, >={Stealth[]}},
  tmd band label/.style={font=\footnotesize\bfseries, anchor=south},
  tmd lane label/.style={font=\scriptsize\itshape, align=flush center, text width=1.45cm},
  tmd event/.style={circle, fill=black, inner sep=0pt, minimum size=4pt},
  tmd msg/.style={->, semithick},
  tmd msg label/.style={font=\scriptsize, align=left, inner sep=1.5pt, anchor=west},
  tmd outcome/.style={draw, rounded corners, align=flush center, fill=#1,
    text width=1.6cm, minimum height=0.7cm, font=\scriptsize},
  tmd outcome/.default=gray!8,
  tmd note/.style={font=\scriptsize\itshape, gray},
  tmd trigger/.style={align=left, font=\scriptsize, anchor=south west, inner sep=1pt},
  tmd bar/.style={draw, fill=white, rounded corners=1pt},
  tmd timer name/.style={font=\scriptsize\itshape, anchor=east, inner sep=1pt},
  tmd expiry label/.style={align=left, font=\scriptsize, inner sep=1pt, anchor=north west},
  tmd expiry/.style={->, dashed},
  tmd clock/.pic={\draw[fill=white, line width=0.5pt] (0,0) circle (0.13cm);
    \draw[line width=0.6pt] (0,0) -- (0,0.09) (0,0) -- (0.065,0);},
}
\newcommand{\tmdframe}[2]{\def\tmdtop{#1}\def\tmdbottom{#2}}
\newcommand{\tmdband}[4][gray!12]{%
  \begin{scope}[on background layer]
    \fill[#1] (#2,\tmdbottom) rectangle (#3,\tmdtop);
  \end{scope}
  \node[tmd band label] at ({(#2+#3)/2},\tmdtop) {#4};}
\newcommand{\tmdlane}[2]{\node[tmd lane label] at (0.75,#1) {#2};}
\newcommand{\tmdlaneline}[3]{\draw[gray] (#2,#1) -- (#3,#1);}
\newcommand{\tmdmsg}[3]{%
  \node[tmd event] at #1 {};
  \draw[tmd msg] #1 -- node[tmd msg label, xshift=4pt] {#3} #2;}
\newcommand{\tmdmsgleft}[3]{%
  \node[tmd event] at #1 {};
  \draw[tmd msg] #1 -- node[tmd msg label, anchor=east, align=right, xshift=-4pt] {#3} #2;}
\newcommand{\tmdtimer}[5]{%
  \draw[tmd bar] (#2,#1-0.14) rectangle (#3,#1+0.14);
  \node[tmd timer name] at (#3-0.05,#1) {#4};
  \pic at (#2,#1) {tmd clock};
  \node[tmd trigger] at (#2+0.05,#1+0.2) {#5};}
\newcommand{\tmdexpiry}[3]{%
  \draw[tmd expiry] (#2,#1-0.14) -- (#2+0.3,#1-0.6)
    node[tmd expiry label] {#3};}
\newcommand{\tmdoutcome}[5][gray!8]{\node[tmd outcome=#1] (#2) at (#3,#4) {#5};}
\newcommand{\gtt}{$>\tfrac{2}{3}$}
\resizebox{\textwidth}{!}{%
\begin{tikzpicture}[tmd]
  \tmdframe{5.95}{-1.45}
  \tmdband{1.6}{4.6}{propose}
  \tmdband[gray!5]{4.6}{9.2}{prevote}
  \tmdband{9.2}{13.9}{precommit}
  \draw[very thick] (1.6,-1.45) -- (1.6,5.95);
  \node[tmd note, anchor=south east, text=black] at (1.55,5.95) {round $r$ starts};
  \tmdlane{5.3}{proposer of round $r$}
  \tmdlane{3.2}{all validators}
  \tmdlane{0.85}{validator (one shown) and its timeouts}
  \tmdlaneline{5.3}{1.6}{13.9}
  \tmdlaneline{3.2}{1.6}{13.9}
  \tmdlaneline{1.3}{1.6}{13.9}
  \node[tmd note, anchor=south east] at (13.85,5.35) {starts no propose timeout; also votes, as one of all validators};
  \node[tmd event] at (1.85,5.3) {};
  \draw[tmd msg] (1.85,5.3) -- node[tmd msg label, xshift=4pt, pos=0.45] {proposal (block)} (3.0,3.2);
  \draw[tmd msg] (1.85,5.3) -- (2.4,1.3);
  \node[tmd event] at (2.4,1.3) {};
  \tmdmsg{(2.85,1.3)}{(3.3,3.2)}{prevote\\ (block or\\ \statevar{nil})}
  \tmdmsgleft{(6.4,3.2)}{(6.9,1.3)}{\gtt{} voting\\ power prevote\\ the block}
  \node[tmd event] at (6.9,1.3) {};
  \tmdmsg{(7.3,1.3)}{(7.75,3.2)}{precommit\\ (block or\\ \statevar{nil})}
  \tmdmsgleft{(11.05,3.2)}{(11.55,1.3)}{\gtt{} voting\\ power precommit\\ the block}
  \node[tmd event] (cm) at (11.55,1.3) {};
  \tmdtimer{0}{1.6}{4.6}{propose timeout}{starts on entering\\ round $r$}
  \tmdtimer{0}{6.9}{9.2}{prevote timeout}{starts once it has prevoted\\ and holds prevotes from\\ \gtt{} voting power, any values}
  \tmdtimer{0}{11.55}{13.9}{precommit timeout}{starts once it holds\\ precommits from \gtt{}\\ voting power, any values}
  \draw[densely dotted, gray] (6.9,1.25) -- (6.9,0.17);
  \draw[densely dotted, gray] (11.55,1.25) -- (11.55,0.17);
  \tmdexpiry{0}{4.6}{\textit{fires first:}\\ prevote \statevar{nil}}
  \tmdexpiry{0}{9.2}{\textit{fires first:}\\ precommit \statevar{nil}}
  \tmdoutcome[gray!35]{dc}{15.45}{1.3}{\textbf{commit} the\\ block; next height}
  \tmdoutcome{nr}{15.45}{-0.75}{\textbf{next round} $r{+}1$,\\ same height}
  \draw[tmd msg] (cm) -- node[above, font=\scriptsize, fill=gray!12, inner sep=1pt, pos=0.3] {commit} (dc.west);
  \draw[tmd expiry] (13.9,-0.14) -- (nr.west);
  \node[tmd expiry label, anchor=north east] at (13.8,-0.3) {\textit{fires first:}\\ next round};
  \draw[->, gray] (1.6,-1.7) -- (16.45,-1.7);
  \node[tmd note, anchor=north east] at (16.45,-1.75) {time in round $r$ of height $h$};
\end{tikzpicture}%
}
\caption{One round~$r$ of a height~$h$ in the baseline at a high level, with time running
left to right. The lanes are the proposer of round~$r$, all validators, and one validator.
Slanted arrows are messages, and the proposal is broadcast to every validator. Solid arrows
are the path when every awaited message arrives in time, on which the validator commits once
precommits from more than two-thirds of the voting power name the block. Each white bar under
the validator's lane is a timeout, which runs from its start trigger to its expiry at the
right end of the bar. The clock marks the start, the label above the bar names the trigger,
and a dotted tick ties the start to the receipt that triggers it. A dashed branch shows what
the validator does when that timeout fires first. The figure follows the description in
Section~\ref{subsec:core-mechanics}.}
\Description{A timing diagram of one round of a height, with time running left to right through
three shaded bands labeled propose, prevote, and precommit. Three horizontal lanes stand for the
proposer, all validators, and one validator. Slanted arrows carry the proposal, the prevotes, and
the precommits between the lanes. Under the last lane, three white bars mark the propose,
prevote, and precommit timeouts. The first two end in dashed branches labeled prevote nil and
precommit nil, and the third ends in a dashed arrow to a box reading next round, same height. A
solid arrow labeled commit leads from the validator's lane to a shaded box reading commit the
block, next height.}
\label{fig:mechanics}
\end{figure*}
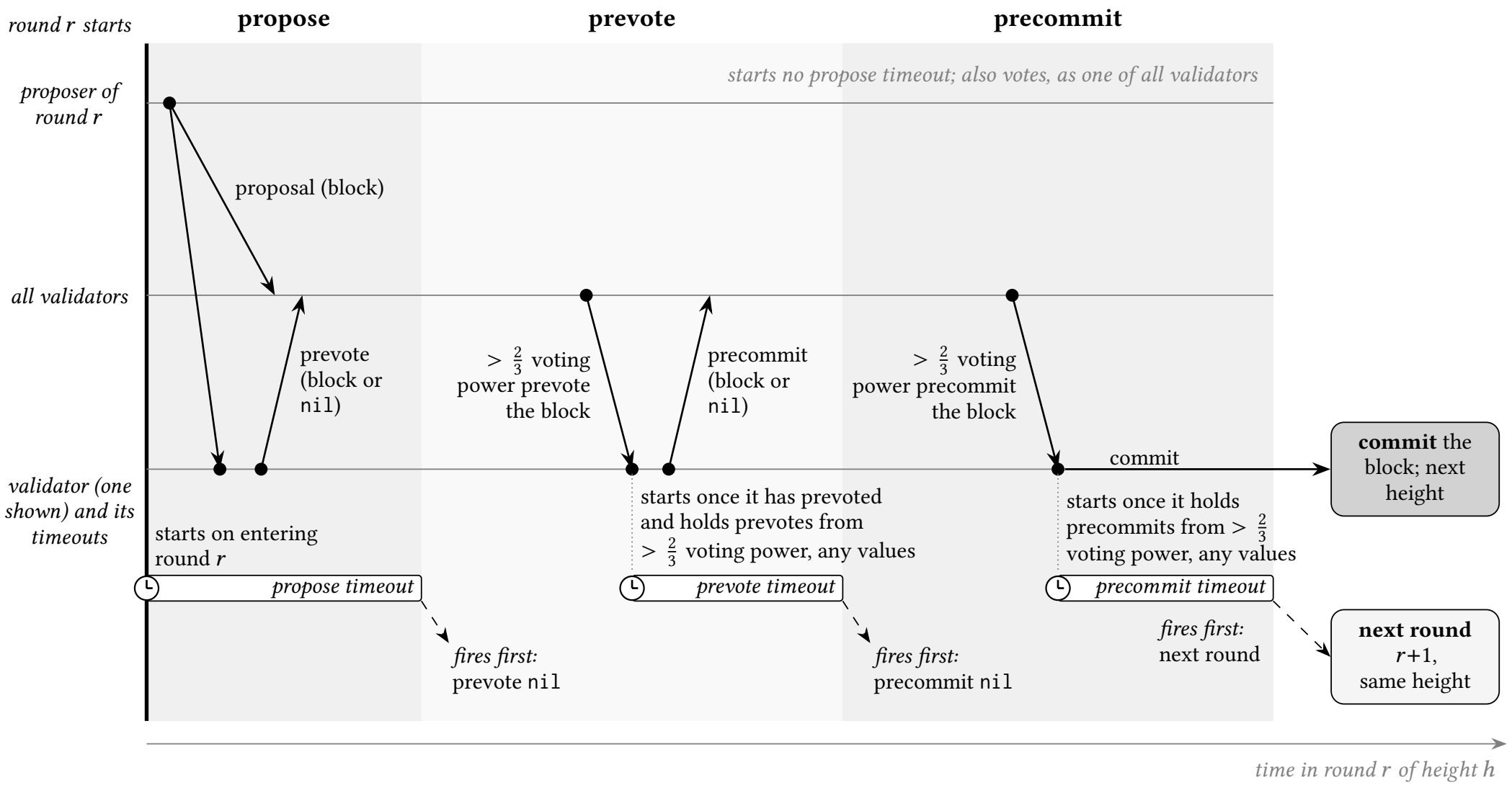

Tendermint grows a blockchain one \emph{height} at a time, and each height decides exactly
one block. Deciding that block may take several numbered \emph{rounds}, each one attempt to
agree on it. The height ends when a round commits a block, and the next height then begins.
Within each round one validator acts as the \emph{proposer}. It is chosen by a
deterministic, voting-power-weighted round-robin that every validator computes identically
for a given height and round, so there is no election and no randomness in this
choice~\cite{buchman2018latest}. The
proposer only suggests a block; the validators decide on it by voting.

The protocol has no \term{vote collector}, a party that gathers the validators' votes and
announces the outcome, so the validators decide among themselves. Each round, as illustrated in Figure~\ref{fig:mechanics}, runs three
steps in order. Each step has a timeout, so that a validator does not wait indefinitely for
messages that may be delayed before GST or that a faulty validator never sends. 
In the \emph{propose} step, the proposer broadcasts a
candidate block, and the other validators start their propose timeout on entering the
round. In the \emph{prevote} step, a validator prevotes for the block if it received the
proposal before that timeout expired, the block is valid, and the validator's lock,
described below, permits it; otherwise it prevotes \statevar{nil}. Whether a block is valid
is decided by an application-specific predicate
(Section~\ref{subsec:abci})~\cite{buchman2018latest,cason2021design}. A prevote means ``I
would back this,'' not a commitment, so prevotes alone never commit a block. Once the
validator has prevoted and has received prevotes from more than two-thirds of the voting
power, for any values and whether or not they agree, it starts its prevote
timeout~\cite{buchman2018latest}. Prevotes carrying more than two-thirds of the total
voting power for the same block form a \term{polka}, which gives the validator evidence
that the block has enough support to become the decision. In the \emph{precommit} step, a
validator that observes a polka precommits that block, which means ``I am ready to commit
this''; if its prevote timeout expires before it observes a polka, it precommits
\statevar{nil}. A validator \emph{commits} the block once it observes a quorum of
precommits for it, carrying more than two-thirds of the total voting power. Its precommit
timeout starts once it has received precommits from more than two-thirds of the voting
power, for any values. If that timeout expires before the validator commits, the validator
moves to the next round of the same height. 

The decision to commit is \emph{local}. Without a vote collector, each validator counts the
votes it sees and commits the moment it observes the precommit quorum, and there is no
separate global agreement step. Within a single round, agreement among the validators holds not because any
party announces it but because every correct validator applies the same quorum rule to the
same votes. That rule is decisive because at most one block can reach a precommit quorum in
a round. By quorum intersection, two quorums would share a correct validator, who casts only one precommit
per round. Hence two quorums for different blocks, in the same round, are impossible. 

A height may run through many rounds before a block commits, and the protocol must stop two
different blocks from committing in different rounds of that height. \term{Locking} does
this. Once a validator precommits a block, it locks on it, recording the block as its
\statevar{locked value} and the round as its \statevar{locked round}. From then on it
prevotes only for that block, unless it sees evidence that the network has moved on, namely a polka
for another block from a round no older than its lock~\cite{buchman2018latest}. For example,
a validator locked on block $A$ in round~3 may prevote block $B$ on a polka for $B$ from
round~5, but not on one from round~2. Locking is safe because, once a block commits, enough
correct validators are locked on it that no other block can gather a polka.

Locks alone can leave validators stuck on different blocks, so that no block gathers a polka.
To avoid this, each validator remembers the latest block it saw win a polka as its
\statevar{valid value}, and the round of that polka as its \statevar{valid round}. When it
becomes proposer, it re-proposes that block and cites that round~\cite{buchman2018latest}.
The citation gives locked validators the evidence they need to prevote the block, so the
height eventually commits after GST.

\subsection{Economic incentive layer}
\label{subsec:incentive-layer}
Tendermint's incentive layer gives validators an economic reason to behave correctly and to take part in consensus. It works through two mechanisms. Under staking, a validator bonds stake to participate and loses part of it, through slashing, if it misbehaves. Under reward distribution, a validator can be paid for helping to commit blocks, and so for following the protocol.

\subsubsection{Staking, bonding and slashing}
The bonded stake that sets a validator's voting power is also what penalties act on. Upon a
provable misbehavior, above all double-signing, part of the misbehaving validator's bonded
stake is destroyed. Double-signing is signing two prevotes, or two precommits, of the same height
and round, either for two different blocks or for a block and \statevar{nil}~\cite{cometbft}.
The penalty backs accountable
safety for equivocation (Section~\ref{subsec:accountability})~\cite{buchman2016tendermint}.

\subsubsection{Reward distribution}
\label{subsec:reward-distribution}
Where slashing penalizes misbehavior, reward distribution pays validators for helping to commit
blocks. In the baseline, the transaction fees of block $H$ are divided among the validators whose signatures for that block are included in block $H{+}1$,
in proportion to their voting power~\cite{kwon2014tendermint}. In production, the application
decides which validators receive a block's transaction fees and any newly issued tokens, and in
what shares. The Cosmos~SDK makes this decision in its application layer rather than in the
consensus core~\cite{cosmossdk}. A deployment that follows the baseline's rule pays for a block
only the validators that helped commit it.
Under this rule, a validator's precommit is
its evidence of having done so, since a block commits only on precommits carrying more than
two-thirds of the voting power. Leaving a validator's precommit out of a block can therefore cost
it financially, either as a lost reward under such a rule or as a missed block toward a penalty
for downtime. The baseline sketches such a penalty, which unbonds a validator that has not signed
for a set number of consecutive blocks and may charge it a small fee in proportion to its bonded
stake~\cite{kwon2014tendermint}. The Cosmos~SDK implements a variant over a sliding window of
recent blocks. A validator whose precommits appear in fewer than a minimum fraction of that
window is removed from the validator set for at least a set period and can lose a fraction of its
bonded stake~\cite{cosmossdk}.

Each block carries a \term{participation record}, which lists the validators that committed
its predecessor and whose precommits its proposer includes. Block $H{+}1$ holds this record as the precommits that committed block $H$,
in a field named \statevar{LastCommit}~\cite{buchman2016tendermint,cometbft}. The proposer of block $H{+}1$
chooses which of these precommits to include, provided the included precommits carry more
than two-thirds of the voting power. A \term{reward rule} then determines what each validator
listed in \statevar{LastCommit} is paid. Under the baseline's rule, these are the validators
that share block $H$'s transaction fees~\cite{kwon2014tendermint}.

A validator commits a block as soon as it holds precommits for it from more than two-thirds
of the voting power, so some precommits for that block may reach the validator afterwards. To give
these late precommits a chance to be included in \statevar{LastCommit}, a validator does not
start the next height at once. It waits a little longer to collect as many of the remaining precommits
as possible~\cite{kwon2014tendermint,cometbft}. This wait is the \term{commit timeout}. Under its
default setting every height waits for it, so it adds the same delay to each block and lowers
throughput, as Section~\ref{subsec:conditional-measured} explains.

\subsection{Gossip layer}
\label{subsec:networking-layer}
The gossip layer is the peer-to-peer \emph{overlay}, the network of direct connections
that each node\footnote{A node is either a validator, which signs votes, or a non-validator node, which follows consensus and processes transactions but does not sign votes~\cite{buchman2016tendermint}.} keeps to a subset of the other nodes, its \emph{peers}, on which the baseline's model assumes the
gossip communication property~\cite{buchman2016tendermint,buchman2018latest}. Because no node
is connected to every other, proposals and votes propagate by gossip over that overlay,
relayed hop by hop rather than sent point-to-point to every node. This
form of dissemination is a core part of the protocol and is common to its implementations. In
CometBFT, a node tracks each peer's height, round, and step and which votes and parts of the
proposed block the peer holds, and sends it only the missing ones~\cite{cometbft}. Pending transactions, unlike proposals and votes, travel over the
same overlay but through the mempool's own, simpler relay (Section~\ref{subsec:mempool}).

\subsection{Separating consensus from the application}
\label{subsec:abci}
The Tendermint protocol separates agreement on the order of blocks from their execution. The validators
agree on each block, and the application, the deterministic state machine that they replicate,
then executes its transactions~\cite{buchman2018latest}. The application also decides which
blocks are valid through an application-specific predicate~\cite{buchman2018latest}.

The engines realize this separation through an interface across which the consensus engine and the
application communicate, namely, \abci{}~\cite{buchman2016tendermint,cason2021design,buchman2022revisiting}.
The protocol assumes that each validator holds an amount of voting power, which may be zero~\cite{buchman2018latest}.
The application supplies the validator set and the voting power of each
validator through \abci{}, and after executing a block it returns any changes to
the validator set~\cite{buchman2016tendermint,buchman2022revisiting}.

\subsection{The mempool and proposal construction}
\label{subsec:mempool}
The transactions that a proposer puts in its block come from its \term{mempool}, the pool of
pending transactions that a node holds before they are
proposed~\cite{buchman2016tendermint,cason2021design}. A node admits a transaction only if the
application accepts it through \checktx, the \abci{} call that checks whether the transaction
would be valid against the latest state~\cite{buchman2016tendermint}. The node holds its
admitted transactions in the order in which it admitted them. Its relay forwards them
best-effort to every peer except those it received them
from~\cite{buchman2016tendermint,cason2021design}. To build a block, the
proposer takes the longest prefix of that order that fits the block's limits, such as its
maximum size in bytes~\cite{cason2021design}.\footnote{In CometBFT, the proposer's application
can also reorder, remove, or add transactions before the proposer sends its
proposal~\cite{cometbft}.} A pending transaction therefore enters a block
only through a proposer that holds it. Once a block commits, each node removes the committed
transactions from its mempool and re-checks the rest through \checktx, since the new state may
have made some of them invalid~\cite{buchman2016tendermint,cason2021design}. Beyond
transactions, a later revision of \abci{} in CometBFT, ABCI++, lets each validator attach
application data, called a \term{vote extension}, to its precommit for a
block~\cite{cometbft}.

\subsection{Guarantees and complexity}
\label{subsec:guarantees}
The protocol's mechanisms yield its guarantees of safety, liveness, accountability, and
instant finality\footnote{Beyond the correctness proofs on which this subsection draws~\cite{buchman2018latest,amoussou2019dissecting}, the safety and liveness of Tendermint have also been studied by other methods. Bounded model checking verifies both properties on a restricted four-validator model, after a fix that the work proposes~\cite{yu2025modelcheck}, and a universal-composability analysis proves termination, the formal liveness property, for a variant of the protocol~\cite{dong2025uc}.},
which this subsection presents together with the protocol's message and verification complexity.

\subsubsection{Safety}
\label{subsec:safety}
Safety means that no two correct validators ever commit conflicting blocks at the
same height. It needs no timing assumption, because the protocol's proof of agreement uses
only the fault bound and quorum intersection. Safety therefore holds even before
GST~\cite{buchman2018latest}. The proof combines two arguments.
Quorum intersection rules out conflicting commits within
one round, and locking extends that exclusion to the later rounds of the
height~\cite{buchman2016tendermint,buchman2018latest,amoussou2019dissecting,amoussou2018correctness}.\footnote{Formally,
the protocol is analyzed against a consensus specification of four properties. Integrity requires
that no correct validator decides twice, Agreement that if a correct validator decides a block,
then eventually every correct validator decides that block, Validity that a decided block
satisfies the application-specific predicate, and Termination that every correct validator
eventually decides~\cite{amoussou2019dissecting}. Integrity, Validity, and the part of Agreement
that forbids correct validators from deciding different blocks are safety properties. The
eventual part of Agreement, like Termination, is a liveness property.}

\subsubsection{Liveness}
\label{subsec:liveness}
Liveness means that the chain keeps committing blocks. It is formalized as
\emph{Termination}, that is, every correct validator eventually decides, and this is guaranteed only
after GST~\cite{amoussou2019dissecting}.
The guarantee rests on two mechanisms, the step
timeouts and the re-proposal of the \statevar{valid value}, together with the gossip
communication property. The step timeouts increase from one round to the next, so an
underestimated message-delay bound does not permanently prevent progress. After GST, the
timeouts grow large enough for the messages of a round to reach every correct validator
before its steps time out~\cite{buchman2018latest,amoussou2018correctness}. The gossip
communication property supplies that delivery, including the prevotes behind each
validator's \statevar{valid value}. The re-proposal therefore eventually yields a block that
all correct validators accept, and the height commits~\cite{buchman2018latest}.
Whether a particular valid transaction commits also depends on whether some proposer includes
it, which Section~\ref{subsec:conditional-measured} treats as censorship resistance.

\subsubsection{Accountability}
\label{subsec:accountability}
\term{Accountability} is the ability to attribute a deviation from the protocol to the
validators responsible for it, from evidence that any party can verify, without ever
attributing a deviation to a correct validator~\cite{haeberlen2007peerreview}. A
\term{deviation} is an action that breaks a rule of the protocol, either sending a message
that the protocol forbids or failing to send one that it requires, and a validator that
deviates is faulty~\cite{haeberlen2007peerreview}. A deviation therefore exists only against a
rule of the protocol. An action that the protocol leaves open, because no rule applies to it, is
not a deviation even when it goes against the protocol's intent, so it is not
attributed~\cite{haeberlen2007peerreview}. Two cases of accountability are named after
the guarantee whose violation must yield the evidence. Under \term{accountable safety}, a
safety violation must yield evidence against some of the validators that caused
it~\cite{buchman2022revisiting}. Under \term{accountable liveness}, a liveness violation must
do so~\cite{lewis2025accountable}.

Tendermint provides accountable safety for \term{equivocation}, the deviation in which a
validator signs conflicting votes in the same round~\cite{buchman2022revisiting}. Safety can
be violated only if more than one-third of the voting power is Byzantine, and a violation then
takes the form of a fork.
The evidence of equivocation comes from quorum intersection. When two conflicting
precommit quorums form in the same round, the validators in their overlap, which hold more
than one-third of the voting power, must have signed precommits for two blocks. Because each
vote carries its signer's signature, any party can verify the two conflicting precommits as
evidence of the deviation.

A fork can also arise from precommit quorums in different rounds of a height, and such a fork
need not involve any equivocation. It can be traced only by collecting, after the fork, the votes
that correct validators saw at that height and matching each vote to the lock or polka that
should justify it~\cite{buchman2016tendermint,buchman2022revisiting}. The deviation behind such a
fork is called \term{amnesia}, in which a locked validator later supports another block without
the polka that would justify it. The signed votes alone do not prove this deviation. The baseline
therefore does not attribute such a fork, and Tendermint does not punish
amnesia~\cite{buchman2022revisiting}.

\subsubsection{Instant finality}
\label{subsec:instant-finality}
Tendermint provides \term{instant finality}, which means that a block is committed and final the moment a
validator observes the precommit quorum, with no probabilistic settlement and no
confirmation depth, and within the fault bound a correct validator never reverts it. Instant finality follows from
the quorum rule and locking established above. When the network is too unreliable for a
quorum to form, the protocol halts rather than forking, trading progress for safety.

\subsubsection{Complexity}
\label{subsec:complexity}
The protocol's message and verification complexity follows from the vote gossip and the
per-vote signatures. Complexity is counted in validators, so in this subsection every validator
is taken to hold the same voting power. Here $n$ denotes the number of validators and $f$
the largest number of Byzantine validators the fault bound allows, which is just under
$n/3$. Because
votes are signed individually and not aggregated, verifying the votes of $n$ validators
takes $O(n)$ signature verifications. Message complexity is counted in the abstract
all-to-all model, where each of the $O(n)$ votes in a round must reach every validator, so
a round costs $O(n^2)$ vote deliveries. After GST, a decision may span up to $O(n)$ rounds in
the worst case, because the correct validator holding the highest lock may not become proposer for
up to $O(n)$ rounds under round-robin rotation. Multiplying $O(n^2)$ per round by $O(n)$
rounds gives a worst-case message complexity of $O(n^3)$ to decide, an order below
PBFT's~\cite{castro1999pbft} $O(n^4)$~\cite{amoussou2019dissecting}.
These are protocol-level costs fixed by the design,
not by any deployment.

\subsection{Measured, conditional, and economic properties}
\label{subsec:conditional-measured}
The guarantees of Section~\ref{subsec:guarantees} are proven invariants, which hold by design
under the fault bound, partial synchrony, and the gossip communication property, independently
of any particular deployment.
Tendermint also has properties that its design does not fix as invariants, and these fall
into three kinds. The first kind is the measured quantities,
throughput, latency, and scalability, each a number reported together with the experimental setup
that produced it. The second kind is the conditional properties, namely proposer
availability, ordering fairness, fairness in reward distribution, and censorship resistance.
Each of them holds or fails against a named adversary or configuration rather than taking a
numerical value. The third kind is economic security, which concerns the proof-of-stake
layer rather than the consensus mechanism itself. Each property can be affected by a deployment factor or condition, a
protocol-level design choice, or both. Section~\ref{subsec:dna-choices} identifies the
protocol-level design choices, and Section~\ref{sec:matrix} determines which surveyed works
strengthen or weaken each property.

Throughput is the rate at which the chain commits work, measured either as transactions
committed per second or as committed data per second, that is, block bytes per second.
Latency is the time that work takes to commit. Block latency is the time a block takes to
commit, and transaction latency is the time from a transaction's submission to its commit.
Because a committed block is final (Section~\ref{subsec:instant-finality}), block latency is
also the time to finality. Both throughput and latency depend on many deployment factors and
conditions, among them the number of validators, the
hardware each validator runs, the network between them, the overlay topology, and the faults that occur.
The overlay topology determines how many peers each node exchanges messages
with~\cite{cason2021design}. Throughput and latency also depend on configuration that a
CometBFT deployment sets, including the maximum block size, the maximum transaction size,
and the consensus timeouts~\cite{cometbft}. Unlike the step timeouts (the propose, prevote,
and precommit timeouts), which delay a height only when a step fails to complete in time, the commit timeout
delays every height under its default setting, so it worsens block latency and lowers
throughput even when no fault occurs. In CometBFT v0.38 and v1.0, it is a configuration
parameter with a default of 1~s, and an operator can let a node skip the wait once it holds the
precommits of every validator~\cite{cometbft}.
In a geo-distributed testbed with 128 validators, each block took on
average about 2.5~s to commit, of which the default commit timeout accounted for about 1~s
and the propose, prevote, and precommit steps together for about
1.5~s~\cite{cason2021design}.\footnote{The
testbed ran Tendermint Core v0.33.8 in its default configuration, with validators spread
evenly across 16 AWS regions, 1~KB transactions, and 20~MB blocks. Clients submitted
transactions in closed loop, that is, each client sent its next transaction only after the
previous one had committed. The measured throughput therefore follows the time a block takes
to commit under the offered load.
The reference workload of 1{,}536 clients was the one the study found to balance throughput
and latency best while saturating the system~\cite{cason2021design}.}

Faults lower throughput further. A height whose scheduled proposer has
crashed commits only after the step timeouts of that proposer's round expire and the turn passes to the next
proposer. Each further silent proposer adds a longer round, because the timeouts grow within
a height~\cite{cason2021design}. In the same geo-distributed testbed with 128 validators,
crashing 42 of them, close to one-third, lowered overall throughput from 438 to 164 transactions per
second, and the latencies of blocks assigned to crashed proposers were dominated by
timeouts~\cite{cason2021design}.

Scalability is how the protocol's performance, gauged for instance by throughput, block
latency, or transaction latency, changes as the validator
count grows. Part of that change comes from the message complexity of
consensus. The protocol's design sets how this complexity grows with the validator count in
the abstract model. How much this complexity lowers throughput and worsens block latency at a
given count, however, depends on the implementation, the hardware, the network, the overlay
topology, and the configuration. In a geo-distributed testbed study~\cite{cason2021design},
an eightfold increase in the validator count, from 16 to 128 validators, reduced throughput
by 18\%, from 535 to 438 transactions per second. The study describes the accompanying worsening
of average block latency, by 18\%, from 2.14 to 2.53~s, as somewhat expected from the quadratic
message complexity of consensus~\cite{cason2021design}.
A
comparative benchmark supports the role of message complexity and reports that,
even with sufficient network bandwidth, the CPU cost of serializing and deserializing
messages limited throughput as the validator count grew~\cite{alqahtani2021bottlenecks}.\footnote{The benchmark reimplemented several
BFT protocols, Tendermint among them, in a common framework and ran them on 4 to 20 two-vCPU
cloud instances over a LAN and a four-region WAN.}
Not all of the worsening of latency comes from consensus. Average transaction latency worsened by 27\%, against 18\% for
block latency. The testbed study attributes this excess to longer delays in propagating
transactions to all validators through the mempool as the validator count
grew~\cite{cason2021design}.

Proposer availability is the resistance of the chain's progress to network-level targeting
of its proposers, that is, to an external network adversary that silences specific honest
proposers by denial-of-service or by eclipse. It differs from liveness, which is proved
under the baseline's model, where the adversary controls
validators holding less than one-third of the voting power rather than targeting honest ones.
An external adversary can silence the proposer of a round to delay the
chain~\cite{bano2019sok}, or silence each proposer in turn to halt
it~\cite{gelashvili2022jolteon}. Each
round has a single proposer, so a silenced proposer stalls its round until the round's step
timeouts expire. The round-robin proposer rule lets the adversary pick its target in
advance, because the rule is deterministic and the proposer schedule is therefore known well
before each turn~\cite{buchman2022revisiting}. Deployments make a targeted proposer
harder to silence through the network topology, without changing the protocol. Each validator is
placed behind multiple \term{sentry} processes, each
acting as a proxy between the validator and the rest of the network and shielding the
validator from inbound traffic and denial-of-service attacks~\cite{buchman2022revisiting}.

Ordering fairness is fairness in the
committed order of transactions. One notion of it is that the order should
reflect the order in which transactions were received rather than being
rearranged for profit. A distributed system, however, has no single global receipt
order against which to measure this notion. Whether ordering fairness holds rests on the
mempool and on the proposal-construction logic a deployment runs.
Resistance to front-running, the reordering that inserts a
transaction ahead of a known pending one for profit, is one instance of this
property.

\emph{Fairness in reward distribution} holds if every validator that took part in
committing a block is recorded as a participant and is rewarded. It therefore holds only under
three conditions, on the commit timeout, the participation record \statevar{LastCommit}, and the
reward rule. The commit timeout must be long enough for late precommits to
arrive, the next proposer must record them all, and the reward rule must pay every recorded
participant. The first two conditions face two threats, one accidental and one deliberate.
The accidental threat is a missed precommit. The next proposer can include only the
precommits that have reached it when it proposes, so the precommit of a slow validator may be
left out. The commit timeout guards against this threat. The deliberate threat is a
proposer that omits a precommit on purpose. Such an omission is argued not to pay, because a
validator whose precommit is omitted can retaliate~\cite{kwon2014tendermint}. When the
retaliating validator's turn to propose comes, it can omit the proposer's precommit. If each of
the two proposes once and the retaliating validator holds a small share of the voting power, the
proposer gains only when its block carries far larger fees than the retaliating validator's
block~\cite{kwon2014tendermint}.

Censorship resistance holds when no coalition within the fault bound can
keep a valid transaction, or the vote of a correct validator, out of the chain
indefinitely. For a transaction, delay is tolerated, so censorship resistance requires only
that the transaction is eventually included.
Whether censorship resistance holds rests on the deployment's proposer and mempool
configuration and on how much voting power the validators willing to censor hold. A single
censoring proposer can only
delay a transaction. It can omit the transaction on its turn, but
the round-robin rotation, which also determines proposer availability, passes the turn
to other validators, and the mempool
has relayed the transaction to them, so an honest proposer
that holds the transaction includes it. The share of voting power held by censoring
validators therefore sets how long the delay lasts. Exclusion becomes indefinite
only when the censoring validators hold at least one-third of the voting power,
beyond the baseline's fault bound. They can then withhold their votes from
every block that contains the transaction and so prevent such a block from gathering
the more-than-two-thirds quorum it needs to commit.

The baseline does not provide censorship resistance for votes. A correct validator's precommit enters the chain
only through the participation record of the next block, its \statevar{LastCommit}, from
which that block's proposer may leave out any precommit beyond the quorum.
Rotation does not undo this omission, as it does
for a transaction, because each block records only the precommits for its
predecessor~\cite{buchman2016tendermint}. A single proposer therefore suffices to keep a
correct validator's vote out of the chain, since no later proposer can add the omitted
precommit.

Economic security is the security of the proof-of-stake layer against
attacks specific to staking, and the consensus core provides it only in part. The core
supplies the equivocation evidence that triggers a penalty, while the application sets the
parameters, such as the bonded stake at risk, the \statevar{unbonding period}, and the
slashing penalties for equivocation and for downtime~\cite{cosmossdk}. One attack specific
to staking is the \term{long-range attack}. In this attack, validators that held more than
two-thirds of the voting power at some past height first unbond and spend their stake. They
then sign a fork of the chain that branches from that height. Slashing does not punish the
fork, because equivocation evidence destroys only stake that is still
bonded~\cite{kwon2014tendermint}.

\section{Design choices and the works that renegotiate them}
\label{sec:design-dna}

We treat Tendermint as a set of design choices that another protocol can make differently yet
still achieve the same safety and liveness guarantees under the same fault and timing
model~\cite{yin2019hotstuff,buchman2022revisiting}. HotStuff is one such protocol. It uses a
leader-based approach, in which the leader collects the votes of each step into a quorum
certificate~\cite{yin2019hotstuff}. The validators commit the proposal once the leader's
certificate for the last step reaches them~\cite{yin2019hotstuff}. Tendermint, by contrast, has
no vote collector and instead uses a gossip overlay to disseminate every vote to every correct
validator~\cite{buchman2018latest}. Each validator commits a block by counting the votes it
receives locally~\cite{buchman2016tendermint,buchman2018latest}.
How votes reach the validators is thus a decision on which the two protocols
differ. We call each decision of this kind a \emph{design choice}. In Section~\ref{subsec:dna-choices}, we first identify Tendermint's design choices and the properties and
guarantees that they provide, individually or in conjunction with other choices.
We then classify the works of Tendermint's lineage by the design
choice that each work \emph{renegotiates} and the properties that the renegotiation \emph{moves}.
A work renegotiates a design choice when it modifies or replaces it, much as a design choice in Bedrock transforms one protocol into another~\cite{amiri2024bedrock}, and it moves a property when it strengthens or weakens that property.
Through this lens, we summarize each work in Section~\ref{sec:matrix} and provide a broader perspective on the implications of the classification in Section~\ref{subsec:map-shows}.

\subsection{Tendermint's design choices and what each provides}
\label{subsec:dna-choices}

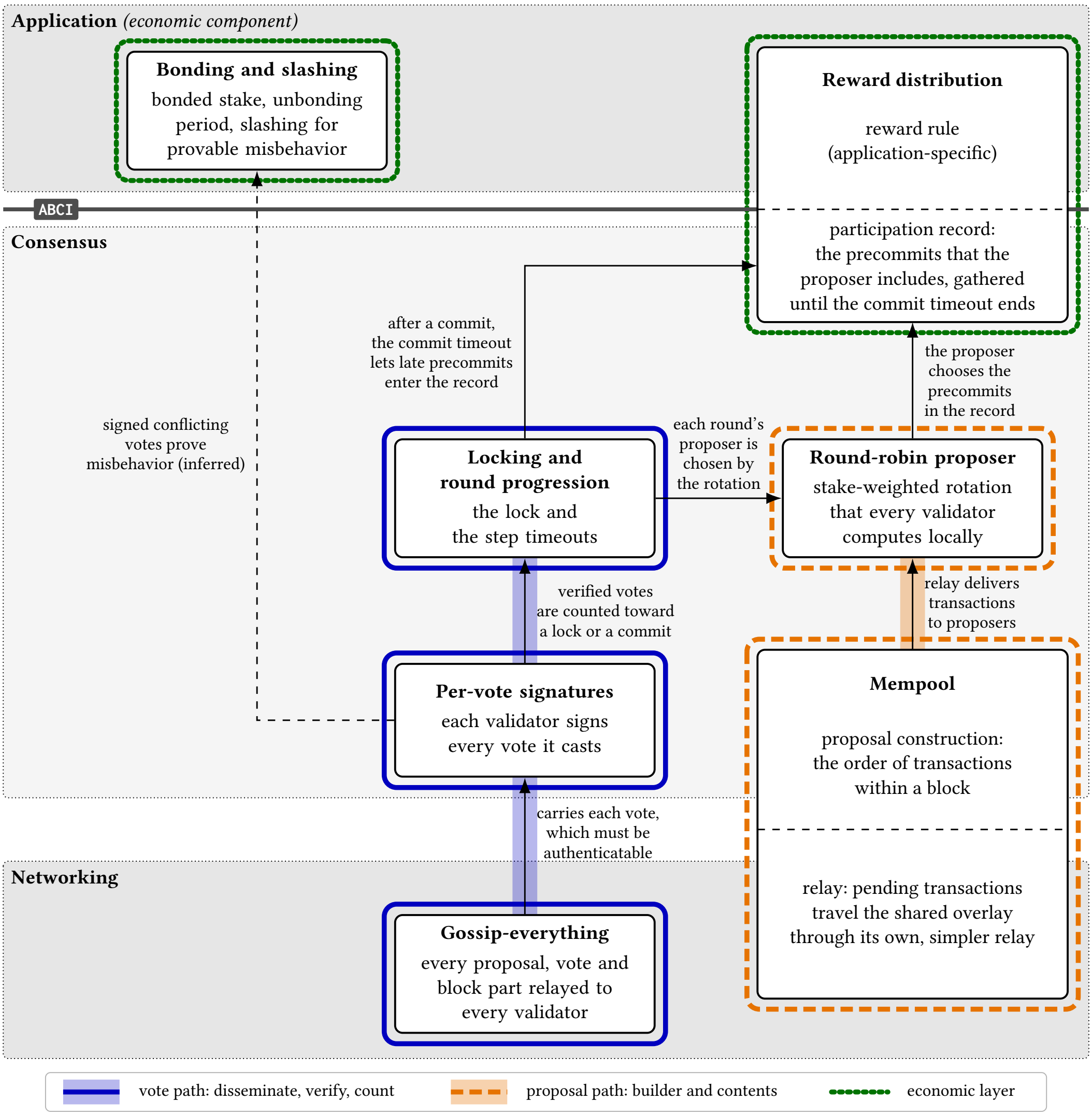
\begin{figure*}[t]
\centering
\colorlet{dcvote}{blue!75!black}
\colorlet{dcprop}{orange!90!black}
\colorlet{dcecon}{green!45!black}
\tikzset{
  dc/.style={font=\footnotesize, >={Latex[length=2.2mm]}},
  dc vote ribbon/.style={line width=10pt, dcvote, opacity=0.28, line cap=round, line join=round},
  dc prop ribbon/.style={line width=10pt, dcprop, opacity=0.32, line cap=round, line join=round},
  dc vote halo/.style={draw=dcvote, line width=2pt, rounded corners=5pt, inner sep=4pt},
  dc prop halo/.style={draw=dcprop, line width=2pt, rounded corners=5pt, inner sep=4pt,
    dash pattern=on 5pt off 2.5pt},
  dc econ halo/.style={draw=dcecon, line width=2pt, rounded corners=5pt, inner sep=4pt,
    dash pattern=on 1.2pt off 2pt, line cap=round},
  dc choice/.style={draw, thick, rounded corners=3pt, fill=white, align=center,
    text width=34mm, inner sep=4pt, font=\footnotesize, minimum height=16mm},
  dc layer/.style={draw, densely dotted, fill=black!#1, rounded corners=2pt},
  dc layer/.default=4,
  dc tab/.style={anchor=north west, font=\footnotesize\bfseries, inner sep=3pt, align=left},
  dc rel/.style={->, semithick},
  dc inferred/.style={->, semithick, dashed},
  dc label/.style={inner sep=1.5pt, font=\scriptsize, align=center},
}
\newcommand{\dcchoice}[5]{%
  \node[dc choice] (#1) at #2 {\textbf{#3}\\[2pt] #5};}
\newcommand{\dclayer}[6]{%
  \begin{scope}[on background layer]
    \draw[dc layer=#1] (#2,#3) rectangle (#4,#5);
  \end{scope}
  \node[dc tab] at (#2,#3) {#6};}
\resizebox{\textwidth}{!}{%
\begin{tikzpicture}[dc]
  \dclayer{10}{0}{0.55}{15.4}{-2.1}{Application \normalfont\itshape (economic component)}
  \dclayer{4}{0}{-2.6}{15.4}{-10.7}{Consensus}
  \dclayer{10}{0}{-11.6}{15.4}{-14.4}{Networking}

  \dcchoice{B}{(3.6,-0.95)}{Bonding and slashing}{econ}{bonded stake, unbonding\\ period, slashing for\\ provable misbehavior}
  \dcchoice{L}{(7.4,-6.45)}{Locking and round progression}{core}{the lock and\\ the step timeouts}
  \dcchoice{P}{(12.9,-6.45)}{Round-robin proposer}{proposer}{stake-weighted rotation\\ that every validator\\ computes locally}
  \dcchoice{S}{(7.4,-9.6)}{Per-vote signatures}{identity \& auth.}{each validator signs\\ every vote it casts}
  \dcchoice{G}{(7.4,-13.2)}{Gossip-everything}{gossip}{every proposal, vote and\\ block part relayed to\\ every validator}

  \draw[thick, rounded corners=3pt, fill=white] (10.7,-0.05) rectangle (15.1,-3.95);
  \node[font=\footnotesize, align=center] at (12.9,-0.5) {\textbf{Reward distribution}};
  \node[font=\footnotesize, align=center] at (12.9,-1.4) {reward rule\\ (application-specific)};
  \draw[semithick, dashed] (10.7,-2.35) -- (15.1,-2.35);
  \node[font=\footnotesize, align=center] at (12.9,-3.15) {participation record:\\ the precommits that the\\ proposer includes, gathered\\ until the commit timeout ends};

  \draw[thick, rounded corners=3pt, fill=white] (10.7,-8.6) rectangle (15.1,-13.55);
  \node[font=\footnotesize, align=center] at (12.9,-9.1) {\textbf{Mempool}};
  \node[font=\footnotesize, align=center] at (12.9,-10.2) {proposal construction:\\ the order of transactions\\ within a block};
  \draw[semithick, dashed] (10.7,-11.15) -- (15.1,-11.15);
  \node[font=\footnotesize, align=center] at (12.9,-12.35) {relay: pending transactions\\ travel the shared overlay\\ through its own, simpler relay};

  \begin{scope}[on background layer]
    \draw[dc vote ribbon] (G.center) -- (S.center) -- (L.center);
    \node[dc vote halo, fit=(G)] {};
    \node[dc vote halo, fit=(S)] {};
    \node[dc vote halo, fit=(L)] {};
    \draw[dc prop ribbon] (12.9,-10.6) -- (P.center);
    \draw[dc prop halo] (10.55,-8.45) rectangle (15.25,-13.7);
    \node[dc prop halo, fit=(P)] {};
    \node[dc econ halo, fit=(B)] {};
    \draw[dc econ halo] (10.55,0.1) rectangle (15.25,-4.1);
    \draw[black!70, line width=1.6pt] (0,-2.35) -- (15.4,-2.35);
  \end{scope}
  \node[fill=black!70, text=white, font=\scriptsize\bfseries, inner sep=2pt,
    rounded corners=1pt] at (0.75,-2.35) {\abci{}};
  \draw[gray!60, rounded corners=2pt] (0.6,-14.6) rectangle (14.8,-15.15);
  \draw[dc vote ribbon, line cap=butt] (0.85,-14.875) -- (1.65,-14.875);
  \draw[dcvote, line width=2pt] (0.85,-14.875) -- (1.65,-14.875);
  \node[anchor=west, font=\scriptsize] at (1.8,-14.875) {vote path: disseminate, verify, count};
  \draw[dc prop ribbon, line cap=butt] (6.35,-14.875) -- (7.15,-14.875);
  \draw[dcprop, line width=2pt, dash pattern=on 5pt off 2.5pt] (6.35,-14.875) -- (7.15,-14.875);
  \node[anchor=west, font=\scriptsize] at (7.3,-14.875) {proposal path: builder and contents};
  \draw[dcecon, line width=2pt, dash pattern=on 1.2pt off 2pt, line cap=round] (11.75,-14.875) -- (12.55,-14.875);
  \node[anchor=west, font=\scriptsize] at (12.7,-14.875) {economic layer};

  \draw[dc rel] (G) -- node[dc label, right=3pt, anchor=west, pos=0.6] {carries each vote,\\ which must be\\ authenticatable} (S);
  \draw[dc rel] (S) -- node[dc label, right=3pt, anchor=west] {verified votes\\ are counted toward\\ a lock or a commit} (L);
  \draw[dc rel] (L) -- node[dc label, above=2pt] {each round's\\ proposer is\\ chosen by\\ the rotation} (P);
  \draw[dc rel] (12.9,-8.6) -- node[dc label, right=3pt, anchor=west] {relay delivers\\ transactions\\ to proposers} (P.south);
  \draw[dc rel] (P.north) -- node[dc label, right=3pt, anchor=west, pos=0.5] {the proposer\\ chooses the\\ precommits\\ in the record} (12.9,-3.95);
  \draw[dc rel] (L.north) |- node[dc label, pos=0.25, left=3pt, anchor=east] {after a commit,\\ the commit timeout\\ lets late precommits\\ enter the record} (10.7,-3.15);
  \draw[dc inferred] (S.west) -| node[dc label, pos=0.75, left=3pt, anchor=east] {signed conflicting\\ votes prove\\ misbehavior (inferred)} (B.south);
\end{tikzpicture}%
}
\caption{Tendermint's design choices at a high level, in three layers. The application on top
holds the economic component above the \abci{} boundary, the consensus engine lies in the
middle, and networking lies at the bottom. The mempool spans the networking and consensus
layers, because its relay shares the gossip overlay while its ordering feeds the proposal.
Reward distribution spans the consensus and application layers, because consensus builds the
participation record while the application applies the reward rule.
Each rounded box is a design choice, with what it covers below its name. The consensus half of
the reward distribution box includes the commit timeout, which lets late precommits still enter
the record.
The highlights mark the vote
path, the proposal path, and the economic layer, along which the design choices are
identified. Solid arrows are relations between choices that
Sections~\ref{sec:background} and~\ref{subsec:dna-choices} state. The
dashed arrow is inferred from two of their statements, that signed conflicting votes prove
misbehavior and that slashing penalizes provable misbehavior.}
\Description{A diagram of three stacked horizontal layers labeled Application (economic
component), Consensus, and Networking, with a thick line labeled ABCI between the top two.
Rounded boxes name the design choices. Bonding and slashing lies in the application layer;
locking and round progression, round-robin proposer, and per-vote signatures lie in the consensus
layer; and gossip-everything lies in the networking layer. A reward-distribution box spans the
application and consensus layers, and a mempool box spans the consensus and networking layers.
Colored ribbons and outlines highlight the vote path, the proposal path, and the economic layer,
as a legend at the bottom shows. Labeled arrows connect the boxes, and one of them is dashed.}
\label{fig:architecture}
\end{figure*}

We regard the baseline as a set of design choices, each a decision through which the baseline
provides a property and which another protocol could make differently under the same fault
and timing model. Under this definition, the parameters that a deployment sets, such as the maximum block
size and the timeout values, are not design choices, because changing a value leaves the
decision that uses it in place. We
examine Tendermint from three angles, the vote path, the proposal path, and the economic
layer, and identify its design choices along each, as Figure~\ref{fig:architecture} shows. Along the vote path lie
gossip-everything, per-vote signatures, and locking and round progression. These choices govern
how a vote is disseminated from the validator that casts it, how its origin is verified, and
how it is counted toward a lock or a commit~\cite{buchman2016tendermint,buchman2018latest}, and
through them the baseline provides its guarantees of safety, liveness, and accountable safety.
Along the proposal path lie the round-robin proposer and the mempool, which determine
which validator builds the block that is
voted on and which transactions that block
contains. The economic layer comprises bonding and
slashing and reward distribution. These choices govern properties that lie outside the scope
of the fault and timing model, namely economic security and fairness in reward distribution.

\subsubsection{Gossip-everything}
\label{subsec:choice-gossip}
The gossip layer arises from the absence of a central vote collector.
Hence, each validator must count the votes
itself, so every vote has to reach every validator, at a cost of $O(n^2)$ vote deliveries per
round. Under gossip-everything, each validator relays every proposal, vote, and block part it
receives to its peers over the overlay~\cite{buchman2016tendermint,buchman2018latest}.
Gossip-everything is the choice through which the baseline's model assumes the gossip
communication property without a direct link between every pair of validators. Relaying can deliver the same message to a validator more than once, which spends
bandwidth on redundant copies; the baseline limits such duplicates by having each node track what its peers already
hold~\cite{buchman2016tendermint}.
A gossip layer that filters what a validator relays, or
that combines votes into fewer messages, lowers the number of messages and can thereby raise
throughput and scalability and improve latency. The liveness guarantee also relies on this layer to satisfy the gossip communication property.
Hence, any renegotiation of this layer must preserve that property to maintain liveness, or be
carefully analyzed for its impact on liveness.

\subsubsection{Per-vote signatures}
\label{subsec:choice-signatures}
Votes traveling over the gossip layer need to be individually
authenticatable, so that any node that receives a vote in the overlay can verify its origin~\cite{buchman2018latest}.
This property is called \emph{transferable authentication}. Pairwise message authentication
does not provide it, which is why Tendermint obtains it by having each validator sign every
vote it casts with its own digital signature. Digital signatures also enable
accountability, since two conflicting votes from one validator, each carrying its signature,
prove that validator's misbehavior to any party, and accountable safety for equivocation rests
on such proof~\cite{buchman2022revisiting}. Checking the signed votes of $n$ validators costs
$O(n)$ signature verifications, a computational overhead that grows with the validator set. A
new scheme can move scalability, because signature schemes such as BLS are one way to reduce
message complexity~\cite{buchman2018latest}. A new signature scheme can also move
accountability, in either direction. A threshold signature, for instance, weakens it, because
it combines the votes of a step into one signature~\cite{yin2019hotstuff}. That signature is
checked once against a single public key, so it no longer shows who signed.

\subsubsection{Locking and round progression}
\label{subsec:choice-locking}
This design choice covers the locking logic, namely
the lock and the re-proposed \statevar{valid value}, together with the consensus timeout mechanism, namely the
step timeouts that drive round progression~\cite{buchman2018latest}. In the baseline, the lock
provides safety across the rounds of a height, and the timeouts, together with the re-proposed
\statevar{valid value}, provide liveness, meaning that every height terminates after
GST~\cite{buchman2018latest}. Together, the locking logic and the
timeouts let a height recover from a round that does not decide with no messages beyond the votes. Each validator
enters the next round when its own timeouts expire, and the new proposer re-proposes its valid
value without first collecting the locks of other
validators~\cite{buchman2018latest,buchman2022revisiting}. We call this \emph{message-free recovery}.\footnote{Recovery in PBFT and HotStuff, which make
the same synchrony assumptions, is not message-free. In both, a new leader first collects
messages from other validators, each carrying the signatures of a quorum that justify its
sender's lock~\cite{buchman2022revisiting}.}
Message-free recovery costs Tendermint \term{optimistic responsiveness}, the property that
after GST a correct proposer makes progress at the speed of actual message delays, even when it
takes over after a round that did not decide~\cite{yin2019hotstuff,buchman2022revisiting}.
The cost arises because a round that does not decide ends only after at least one correct
validator waits out the round's precommit timeout, so the round lasts longer than the actual message
delays~\cite{buchman2018latest,buchman2022revisiting}.
Because the
lock and the timeouts provide the baseline's safety and liveness, a work that alters the
locking logic or the consensus timeouts can move these guarantees directly. 
Such a work can also move throughput and latency, since the locking logic and the timeouts determine how long each height takes to commit.

\subsubsection{Round-robin proposer}
\label{subsec:choice-proposer}
This design choice covers the proposer-selection rule, a rotation weighted by voting power that every
validator computes locally. The rule chooses proposers in proportion to stake, and since it is
deterministic, it needs no message exchange between validators and keeps recovery
message-free~\cite{buchman2016tendermint,buchman2022revisiting}. Because the turn keeps
passing, no single proposer decides for long what enters the chain, which supports transaction censorship
resistance. However, the price of determinism is that every future proposer is known in advance, which
impacts proposer availability. A different proposer-selection rule can shift these trade-offs, and so move
proposer availability and transaction censorship resistance.

\subsubsection{Mempool}
\label{subsec:choice-mempool}
This design choice covers the mempool logic, including the mempool's relay,
and proposal construction, namely the order of transactions within a block. In CometBFT, the
mempool and the application settle that order together across the \abci{} boundary~\cite{cometbft}.\footnote{CometBFT's own
\texttt{nop} mempool goes further and leaves the storage, dissemination, and proposing of
transactions to the application~\cite{cometbft}.
Narwhal, outside the Tendermint family, separates the reliable dissemination of transactions
from their ordering~\cite{danezis2022narwhal}.}
That order decides ordering fairness. In the baseline, the proposer takes transactions in the
order in which its own mempool admitted them, and each node relays them best-effort to its
peers~\cite{buchman2016tendermint,cason2021design}. The relay supports censorship resistance,
because it lets a later proposer hold a transaction that an earlier one omitted. Without the
relay, a transaction submitted to a single validator that ignores it would never reach any
other validator, and the rotation of proposers could not bring it into a block. The mempool
also affects throughput, because a block can carry only the transactions that have reached its
proposer by the time it proposes. Throughput therefore depends on how quickly the relay
delivers transactions to proposers~\cite{cason2021design}. Because this design choice decides
ordering fairness, supports censorship resistance, and affects throughput, a work that alters
the mempool or proposal construction can move all three properties.

\subsubsection{Bonding and slashing}
\label{subsec:choice-bonding}
This design choice covers
the bonded stake, the unbonding period, and the slashing penalty for provable misbehavior~\cite{buchman2016tendermint}. In
the baseline, bonding provides Sybil resistance, because a party that spreads its stake over
many identities gains no voting power by doing so. Slashing makes equivocation costly, because
a validator that equivocates loses part of its bonded stake~\cite{buchman2016tendermint}.
Both contribute to economic security. Nevertheless, a validator's stake
remains slashable only until its unbonding period ends~\cite{kwon2014tendermint}. Because bonding and
slashing contribute to economic security and the unbonding period limits that contribution, a
work that alters any of the three can move that property. Such a work can also move
accountability when it makes attributable, and hence slashable, a fault that the baseline
leaves unattributed.

\subsubsection{Reward distribution}
\label{subsec:choice-reward}
This design choice covers the mechanisms by which a validator
can be paid for helping to commit blocks. They are the participation record that every block
must carry (the \statevar{LastCommit} field that records the precommits of the previous height), a reward rule, which varies from one application to another, and the commit
timeout. Under a rule that pays from the participation record, an omitted precommit leaves its validator
unpaid, so the record decides fairness in reward distribution. Fairness therefore depends on
how complete the record is. That completeness depends in turn on the proposer and on the
commit timeout. The proposer of each block chooses which precommits beyond the quorum the
record includes. The commit timeout lets late precommits still enter the
record~\cite{cometbft}. This
gain in completeness comes at a price, because the commit timeout adds the same delay to every
height, so it lowers throughput~\cite{cason2021design}. Because the record decides fairness in reward distribution,
and both the proposer and the commit timeout determine what it holds, a change to the record, to
the rule, or to how the commit timeout waits can move that property.

\subsection{The surveyed works, by design choice}
\label{sec:matrix}

We present the classification in Table~\ref{tab:matrix}. Its two dimensions are the design
choice that each work renegotiates and the properties that its contribution strengthens or weakens. We also identify whether each result in a cell holds under the baseline's
fault and timing model or requires a stronger one, and we mark the latter with $\oplus$.

\begin{table*}[t]
\centering
\caption{The classification. Each row is a design choice,
followed in parentheses by the component in which it lies,
and each column is a property. In the row labels, Gossip
stands for gossip-everything, Signatures for per-vote signatures, and Proposer for the
round-robin proposer. Among the components, identity \& auth.\ stands for identity and
authentication and econ for the economic component, and the label \emph{optional} marks that
the reward rule varies from one application to another. In the column headings, Account.\ stands for
accountability, Scalab.\ for scalability, Prop.-avail.\ for proposer availability, Fair-ord
for ordering fairness, Fair-rew for fairness in reward distribution, Censor.\ for censorship
resistance, and Econ-sec for economic security.}
\label{tab:matrix}
\renewcommand{\arraystretch}{1.5}
\setlength{\tabcolsep}{3pt}
\scriptsize
\resizebox{\textwidth}{!}{%
\begin{tabular}{bL{2.2cm}bL{1.15cm}gL{1.3cm}gL{1.2cm}gL{1.35cm}gL{1.35cm}gL{1.2cm}gL{1.15cm}gL{1.2cm}gL{1.2cm}gL{1.15cm}gL{1.25cm}b}
\hline
\textbf{Choice (component)} & \textbf{Safety} & \textbf{Liveness} & \textbf{Account.} & \textbf{Throughput} & \textbf{Low latency} & \textbf{Scalab.} & \textbf{Prop.-avail.} & \textbf{Fair-ord} & \textbf{Fair-rew} & \textbf{Censor.} & \textbf{Econ-sec} \\
\hline
\arrayrulecolor{gridgray}
\textbf{Gossip} \emph{(gossip)}       & & & $\oplus$ [sync] \cite{xian2019improved} & $\CIRCLE$~\cite{cason2025boosting} & $\CIRCLE$~\cite{cason2025boosting}\newline $\blacktriangledown$ [sync, trusted aux.] \cite{xian2019improved} & $\CIRCLE$~\cite{cason2025boosting} & & & $\oplus$ [sync, trusted aux.] \cite{xian2019improved} & $\oplus$ [sync, trusted aux.] \cite{xian2019improved} & \\ \hline
\textbf{Signatures} \emph{(identity \& auth.)}   & $\oplus$ [TEE] \cite{beltrando2023tendertee} & $\oplus$ [TEE] \cite{beltrando2023tendertee} & $\oplus$ [TEE] \cite{beltrando2023tendertee} & & & & & & & & \\ \hline
\textbf{Locking and round progression} \emph{(core)}        & & $\oplus$ [sync, \mbox{Bitcoin}] \cite{tas2023bitcoin} & $\oplus$ [bounded async] \cite{lewis2025accountable} & $\CIRCLE$~\cite{lei2020chained} & $\blacktriangledown$~\cite{lewis2025accountable} && & & & & \\ \hline
\textbf{Proposer} \emph{(proposer)}   & & & & & & & $\CIRCLE$~\cite{bui2023enhanced} & & & & \\ \hline
\textbf{Mempool} \emph{(mempool)}     & & & & & & & & $\CIRCLE$~\cite{momeni2022fairblock} & & & \\ \hline
\textbf{Bonding and slashing} \emph{(econ)}        & & & $\oplus$ [bounded async] \cite{lewis2025accountable}\newline $\oplus$ [sync, \mbox{Bitcoin}] \cite{tas2022babylon} & & & & & & & $\oplus$ [sync, \mbox{Bitcoin}] \cite{tas2022babylon} & $\oplus$ [sync, \mbox{Bitcoin}] \cite{tas2022babylon, tas2023bitcoin} \\ \hline
\textbf{Reward distribution} \emph{(econ, optional)} & & & & & & & & & $\CIRCLE$~\cite{amoussou2018correctnessext} & & \\
\arrayrulecolor{black}\hline
\end{tabular}%
}
\par\smallskip
\begin{minipage}{\textwidth}
\raggedright\footnotesize
$\CIRCLE$: a work in the research literature renegotiates the row's design choice and moves the
column's property, or analyzes such a renegotiation. A work whose design itself makes the
property worse than in the baseline is marked with $\blacktriangledown$ instead, followed by any
tags as for $\oplus$.\\
$\oplus$~[$\cdot$]: the cell's result holds only under assumptions or components outside the
baseline's fault and timing model, and the bracket names each timing assumption and trusted
component that the result adds. The
tags are a synchronous network [sync], a bound on asynchrony [bounded async], a trusted
execution environment [TEE], an external chain [Bitcoin], and trusted auxiliary validators
[trusted aux.].\\
Blank: no surveyed work moves the column's property by renegotiating the row's design choice.
\end{minipage}
\end{table*}

\subsubsection{Filtering and bundling gossiped votes for throughput at scale}
\label{subsec:semantic-gossip}

Along the vote path, gossip-everything requires every prevote and every precommit to reach every validator, so the number of vote messages that each validator handles grows with the validator set. To make progress, however, consensus needs only a quorum of matching votes for each height, round, and step. Once a node holds such a quorum, any further vote for that step no longer contributes to progress. Forwarding it to peers only adds messages. Semantic Gossip~\cite{cason2025boosting} addresses this message cost by making the gossip layer consensus-aware, as explained below.

In Semantic Gossip, the consensus layer supplies two operations, filtering and bundling, that the gossip layer applies before a node forwards votes. Filtering withholds a vote from a node's peers under two rules, while the node's own consensus still receives it. The first rule withholds a vote that the node has already processed. The second rule, post-quorum suppression, withholds every redundant vote, that is, any vote for a height, round, and step at which the node already holds a quorum. To a peer, a withheld vote looks like ordinary message loss, which cannot violate the baseline's safety. Liveness is argued to hold as well, because every vote still needed for a quorum is forwarded~\cite{cason2025boosting}. Bundling, which Semantic Gossip calls semantic aggregation, packs the votes that are queued for a peer and share a type, height, round, and block into one message, and the node sends that message to the peer. The peer unpacks it into the original votes, each with its own signature. Bundling therefore reduces the number of messages that a peer receives, not the number of votes that it processes.

Relative to CometBFT's relay, which already sends each peer only the votes that the peer lacks (Section~\ref{subsec:networking-layer}), we identify two new elements in Semantic Gossip, post-quorum suppression and bundling. Both change only how votes are relayed, so Semantic Gossip renegotiates gossip-everything. Its evaluation, which comes next, shows that Semantic Gossip improves throughput, latency, and scalability.

Semantic Gossip, with filtering and bundling both enabled, is evaluated against classic gossip, a push gossip layer that treats each vote as opaque, in a Go prototype of Tendermint~\cite{cason2025boosting}. With throughput counted as blocks decided per second, the gain over classic gossip rose from $1.24\times$ at $32$ validators to $3.42\times$ at $128$.\footnote{To raise the message load, the experiments ran several consensus instances concurrently over the same validators. This concurrency is a setup of the test rather than a change to the protocol, and a CometBFT chain decides one block at a time. The gains above were taken at the number of concurrent instances that gave each setup its best throughput relative to latency~\cite{cason2025boosting}.} Median latency at $128$ validators was also $2.67\times$ better than with classic gossip~\cite{cason2025boosting}. At $128$ validators, where the throughput gain was largest, several nodes running classic gossip were at full CPU utilization even under the lightest load tested. Message handling was a main source of that load. Semantic Gossip, by contrast, cut the gossip messages that a node received per decided block by about $80$ percent at $128$ validators. The throughput gain is attributed to this reduction in message handling~\cite{cason2025boosting}.

The growth of the throughput gain and of the message cut with the validator set establishes the scalability effect. The message cut grew from about $35$ percent at $32$ validators to about $80$ percent at $128$. These results show scalability in the validator count, measured at two sizes within a single cluster. The evaluation does not cover scalability across a geographically distributed deployment. By design, Semantic Gossip lowers the number of messages per round, whereas the number of vote deliveries per round stays $O(n^2)$, as in the baseline, because each validator still collects individually signed votes toward its quorum.

\subsubsection{Detecting vote-inclusion censorship by auditing validators' reported vote sets}
\label{subsec:vote-censorship}

Section~\ref{subsec:conditional-measured} shows that the baseline does not provide censorship resistance for votes, because a proposer can leave a correct validator's precommit out of the participation record. A coalition of colluding validators can repeat this omission at every height at which one of its members proposes~\cite{xian2019improved}. Under a reward rule that pays from that record, the excluded validator goes unpaid (Section~\ref{subsec:choice-reward}). The baseline does not attribute this vote-inclusion censorship. Its rules permit the omission (Section~\ref{subsec:reward-distribution}), and a vote left out on purpose looks the same as one lost to the network~\cite{xian2019improved}.

The construction proposed in~\cite{xian2019improved} detects such censorship by adding three message types and an auxiliary-validator role. First, each validator votes on the latest block, block $H$. Every validator that receives a vote must acknowledge the receipt with a \emph{vote reply} signed by itself. Unlike the baseline's local and unsigned tracking of which votes a peer holds (Section~\ref{subsec:networking-layer}), this signature lets the reply serve as evidence against its signer. From these vote replies, each validator forms the set of validators that acknowledged its own vote before a timeout. Coalition members withhold the vote reply to each vote they censor, so that they leave no such evidence. Every correct validator's set therefore lacks the same validators, the coalition. Once block $H{+}1$ appears, carrying or leaving out the votes on block $H$, each validator reports its set to the auxiliary validators in a \emph{validation message}. The auxiliary validators compare the reported sets pairwise by their similarity, $S(T_i,T_j) = |T_i \cap T_j| / |T_i \cup T_j|$, against a fixed threshold.\footnote{The work sets the threshold at $0.99$ for at most $2{,}000$ validators, so that two sets count as similar when fewer than $20$ of their validators differ.} A sharp split of the sets into two groups shows a censorship attack. Block $H{+}1$ then tells the two groups apart. The validators whose votes it left out form the correct group, and the other group is the coalition. By contrast, a single validator hit by a network failure is not treated as a coalition. In a network failure, a validator cannot exchange messages with many other validators, so its own set diverges alone or it is missing from the sets of the others. A validator that cannot reach only a few others is not in a network failure, and its set still counts as similar~\cite{xian2019improved}. The construction penalizes a validator for every network failure it suffers, including the first. Each penalty costs the validator a share of its deposit, which the work suggests setting at 10\% the first time and 20\% the second time, rising with each repetition. As the proposer, it loses its block reward instead~\cite{xian2019improved}. Finally, the auxiliary validators send the correct validators a \emph{validation reply} that names the chain to follow, either the original chain or, after censorship, a fork that drops block $H{+}1$. Validators switch to the named chain automatically.

The proposer of block $H{+}1$ can evade the comparison of reported sets by answering the correct validators' votes with vote replies, so that it appears in their reported sets, while still leaving those votes out of block $H{+}1$. The auxiliary validators catch this evasion by checking block $H{+}1$ against the reported sets, and they act when many validators whose sets contain the proposer have their votes left out. The construction thus adds a rule that the baseline does not have, namely that a proposer must include the votes it acknowledged. The work uses the proposer's own signed vote replies, together with block $H{+}1$, as evidence that the proposer broke this rule, and divides its deposit among the correct validators~\cite{xian2019improved}. The work does not specify what a vote reply signs, so this evidence binds the proposer to the omitted votes only if each reply signs the vote that it acknowledges. With such replies, any party can verify this evidence, so the construction makes the proposer's omission attributable. Because the construction changes how validators exchange and acknowledge votes, it renegotiates gossip-everything and moves censorship resistance and accountability. Protecting the inclusion of votes also protects fairness in reward distribution, because, under a reward rule that pays only included precommits, the votes that a block includes decide which validators are rewarded (Section~\ref{subsec:reward-distribution}).

The construction adds assumptions to the baseline's model. It assumes that, outside periods of network failure, every message arrives within a delay bound $\delta$. Its detection also relies on the coalition reaching a certain scale and on its members behaving alike, each rejecting the same votes, so that their reported sets agree and stand apart from those of correct validators~\cite{xian2019improved}. The work places no upper bound on the coalition's size, so it fixes no fault threshold. Finally, the construction relies on correct auxiliary validators, an assumption the work leaves implicit, since validators follow their validation reply automatically and the work does not consider a faulty auxiliary validator. The work selects the auxiliary validators as the validators that performed best in the previous cycle of blocks~\cite{xian2019improved}.

The work argues informally that no defense against censorship can achieve security, low protocol overhead, and a short time to finality all at once~\cite{xian2019improved}. Its method favors security, identifying a censorship attack automatically, punishing the attackers, and forming a chain of correct validators. This leaves room to reduce the length of its added messages, which the work bounds between $(n+1)/2$ and $n$ times the length of one vote reply plus one validation message, where $n$ is the number of validators~\cite{xian2019improved}. It also leaves room to shorten its time to finality, which the work bounds at $O(n \log n) + 3\delta$, above the baseline's, because the auxiliary validators first compute the set of correct validators~\cite{xian2019improved}.\footnote{The bound is analytical, not a measured latency.}

\subsubsection{Raising the fault bound with trusted hardware}
\label{subsec:tendertee}

The baseline's fault bound limits Byzantine validators to less than one-third of the voting power (Section~\ref{sec:background}).\footnote{With equal voting power, this bound matches the requirement of BFT protocols in general under partial synchrony, at least $3f+1$ replicas to guarantee both safety and liveness with $f$ Byzantine ones~\cite{dwork1988consensus}. Trusted hardware restricts the malicious behavior of replicas, so that safety can be guaranteed with $2f+1$ replicas~\cite{amiri2024bedrock}.} Once they hold more than one-third, safety can be violated, and the result is a fork (Section~\ref{subsec:accountability}). To relax this bound, the construction proposed in~\cite{beltrando2023tendertee} places trusted hardware on every validator and raises the tolerated share of Byzantine validators from below one-third to below one-half. The gain comes from a trusted abstraction that makes the protocol immune to equivocation, a malicious behavior in which a faulty validator sends different messages to different validators. This abstraction is an attested append-only memory (A2M), a small trusted component that each validator runs. The component holds append-only logs in which each slot belongs to one protocol step. Before sending a consensus message, a validator appends it to the log for its step and obtains an attestation. The attestation accompanies the message, so that every receiver can verify it against the sender's public key. Because a slot accepts only one value, a validator cannot obtain attestations for two conflicting messages in the same step. A faulty validator therefore becomes a non-equivocating Byzantine one, not a correct one. It can still stay silent, delay its messages, or withhold them. Non-equivocation alone, however, does not raise the tolerated share. It binds each validator to a single value per protocol step, but a correct validator still cannot forward what it heard as evidence that others can check~\cite{beltrando2023tendertee}. The construction therefore pairs the A2M component with digital signatures, which provide transferable authentication (Section~\ref{subsec:choice-signatures}). Correct validators can then act on forwarded messages. With both the A2M component and digital signatures, the construction runs with a committee of $n = 2f+1$ validators, of which at most $f$ are Byzantine, and decides on a quorum of $f+1$ votes from distinct validators~\cite{beltrando2023tendertee, amoussou2025brb}. Because the construction changes how validators authenticate their votes, it renegotiates per-vote signatures and moves safety, liveness, and accountability.

The construction needs assumptions beyond the baseline's model. It assumes that the A2M component on every validator is trusted~\cite{beltrando2023tendertee}. It also assumes a reliable Byzantine broadcast among validators, unforgeable signatures, and a public-key infrastructure in which each validator is identified by the public key of its A2M component. Finally, the work counts validators rather than voting power, so its bound carries over to voting power only when all validators hold equal voting power.

Under these assumptions, the work establishes its results analytically, by proving agreement and termination for one-shot and repeated consensus~\cite{beltrando2023tendertee}. The gain therefore covers liveness as well as safety. For accountability, the construction prevents equivocation, the deviation that the baseline attributes after a fork (Section~\ref{subsec:accountability}), so that it no longer needs to be attributed. This gain is prevention rather than attribution, and it holds only while the A2M component on every validator stays trusted.

\subsubsection{Pipelining consecutive heights for throughput by deferring each block's commit}
\label{subsec:chained}

In the baseline, a validator enters a new height only after it commits the current block, so each block waits for the voting of the one before it. Throughput is therefore set by how long each height takes to commit, which is determined by the locking logic and the timeouts (Section~\ref{subsec:choice-locking}). Chained Tendermint~\cite{lei2020chained}, a construction inspired by Tendermint and HotStuff~\cite{yin2019hotstuff}, is built in stages, and its first, Parallel Tendermint, uses a pipelining idea, in which each validator enters the next height before it commits the current block. The validator sends its precommit for that block later, once it has seen a polka for it. A new proposal extends the block its proposer last prevoted, which need not yet have a polka~\cite{lei2020chained}. As a result, the voting of one height overlaps with the proposal and execution of the next. This is possible because the construction replaces the rounds of a height with successive heights, so a height need not commit its block before the next one begins. Accordingly, the baseline's locking rule carries over with heights in place of rounds, so that a validator keeps its locked block until it sees a polka for a block at a greater height~\cite{lei2020chained}. Likewise, a height whose block gathers no polka in time plays the role of a failed round. Validators move on to the next height when a timeout expires, and the next proposal may extend an earlier block, so that the chain skips that height~\cite{lei2020chained}. Because the construction indexes the lock by height and changes the condition for entering the next height, it renegotiates locking and round progression, and it improves throughput while deferring each block's commit.

The price of moving on before a commit is a deferred commit. Because a validator can send its precommit for a block after it has moved on to later heights, the block can commit some heights after it is proposed. After GST, with each height's timeouts set to half the delivery bound, the work argues that this delay never exceeds five heights~\cite{lei2020chained}. The work describes the deferral as a loss of instant finality, because in its sense the block of each height must be final before a validator enters the next~\cite{lei2020chained}. What the deferral delays, however, is the commit, not the finality of a committed block. The reason is that the work argues that the baseline's safety argument carries over from rounds to heights, so that once a block at height $h$ has a precommit quorum, no conflicting block obtains a polka at height $h$ or above~\cite{lei2020chained}. Without a polka such a block cannot commit, so under the definition of instant finality in Section~\ref{subsec:instant-finality}, a committed block is final as soon as it commits.

Against this cost, the work argues the throughput gain analytically. In the normal case, where the baseline commits each block in a single round, the overlap removes the time of prevoting and precommitting from each block~\cite{lei2020chained}. When a baseline height fails a round, the baseline must propose and vote again before it commits a block. By contrast, the pipeline votes while it proposes and executes later blocks, so it keeps producing blocks during that time~\cite{lei2020chained}. Not all of these blocks commit, however, because the chain can branch. Branches can arise even when all validators are correct, because a proposal may skip a height whose block gathers no polka in time. Only one branch commits, and a validator that executed blocks on a losing branch must then roll that execution back~\cite{lei2020chained}. Even so, the work states that the pipeline commits more than one block in the time that the baseline spends on one multi-round height~\cite{lei2020chained}.

\subsubsection{Unpredictable proposer selection through a verifiable random function}
\label{subsec:vrf-proposer}

Section~\ref{subsec:conditional-measured} shows that the round-robin proposer affects proposer availability, because its deterministic, voting-power-weighted schedule lets an attacker compute each proposer and hence launch a denial-of-service attack on it before its turn. The construction proposed in~\cite{bui2023enhanced} makes this schedule unpredictable more than one height ahead by seeding proposer selection with the output of an elliptic-curve verifiable random function (VRF). A VRF is a primitive with which the holder of a secret key computes, for a given input, a unique pseudorandom value together with a proof that any party can check against the matching public key. The proposer of each block produces such a value and attaches it, with its proof, to its proposal, and every validator checks the proof before it prevotes for the block. Once the block commits, the value it carries seeds the selection of the proposers at the next height.\footnote{The work defines the selection as a deterministic function of the validator set, the round, and the seed from the previous block, without specifying how the seed is mapped to a particular validator~\cite{bui2023enhanced}.} Since that seed is public, selection remains a deterministic function of public inputs, so each validator still computes the proposer of every round locally, with no extra messages~\cite{bui2023enhanced}, and recovery from a failed round stays message-free. As the seed is computed by the proposer of the block before a height and carried in its proposal, an attacker can learn that height's proposers once that proposal is broadcast, although they are fixed only when the block commits~\cite{bui2023enhanced}. The work renegotiates the round-robin proposer selection mechanics of the baseline in favor of enhancing proposer availability.

The work evaluates the construction in a four-node prototype on a local network~\cite{bui2023enhanced}. Under a coordinated denial-of-service whose attacker followed the round-robin schedule, 92.5\% of 1000 rounds timed out on a Tendermint chain, against 52.8\% on the VRF chain~\cite{bui2023enhanced}.\footnote{On both chains, the evaluation targets proposers with an attack pattern fixed in advance, which leaves room to measure the gain against an attacker that also uses the seed once it is revealed~\cite{bui2023enhanced}.} Fewer rounds timed out on the VRF chain because the attacker could no longer compute each proposer from the round-robin schedule, so the construction improves proposer availability. It does so inside the protocol, as a counterpart to the sentry-node topology that deployments apply below the protocol (Section~\ref{subsec:conditional-measured}). The gain in proposer availability comes with a small overhead, because computing and verifying the VRF adds to the formation of each block. The work describes this overhead as slight and the construction's performance as comparable to that of a Tendermint chain~\cite{bui2023enhanced}. In separate runs at about three transactions per second, the overhead was about 12~ms per node, so that over 10,000 blocks a block took on average about 958~ms to finalize on the VRF chain, against about 946~ms on the Tendermint chain~\cite{bui2023enhanced}. Beyond this small overhead, the work also shows that randomizing the selection keeps it weighted by voting power, as the round-robin rule is. In further runs over about 10,000 blocks, each validator proposed a share of the blocks close to its share of the voting power~\cite{bui2023enhanced}.

Even so, some proposers remain predictable under the construction. Once a block commits, the seed it carries stays fixed for the whole next height, so an attacker can derive every round proposer of that height in advance, including those that take over after a timeout, and target each in turn. The same holds at the start of the chain, because the seed before the first block is a public constant, so the proposers of the first height are known before the chain starts. Apart from targeting, the seed is produced by a single party, the block's proposer, rather than by a distributed randomness beacon in which many parties jointly produce each value, so that proposer might try to bias it. The work considers such bias for the same seed, which the chain also offers to smart contracts as a random value. It argues that the proposer cannot choose the input from which the seed is computed, so it can influence the seed only by abstaining from proposing, which risks a slash of its stake and still does not guarantee a favorable outcome~\cite{bui2023enhanced}.

\subsubsection{Front-running prevention through distributed identity-based encryption}
\label{subsec:fairblock}

In the baseline, the proposer orders a block's transactions as its mempool admitted them (Section~\ref{subsec:mempool}). Before that order is fixed, each pending transaction is relayed from node to node, and each node that relays it can read it. Whoever controls the order, or observes pending transactions in the mempool, can therefore gain maximal extractable value (MEV), the revenue obtained by reordering, inserting, or censoring transactions~\cite{momeni2022fairblock}. Most MEV is extracted through front-running by a party other than the block proposer, such as a bot that watches the public mempool and places its own transaction ahead of a pending one by offering a higher fee where fees set the order~\cite{momeni2022fairblock}. A validator can also front-run by ordering its own transaction first when it proposes, and other parties can reach the proposer's mempool first over a faster network~\cite{momeni2022fairblock}. Because such an insertion reorders transactions for profit, Section~\ref{subsec:conditional-measured} treats resistance to front-running as an instance of ordering fairness. Front-running needs the content of a pending transaction before the order is fixed, so a known defense encrypts transactions, orders the ciphertexts, and decrypts them only afterwards. In prior schemes of this kind, each transaction is threshold-decrypted separately, so a committee of size $S$ exchanges $N \times S$ messages for a block of $N$ transactions~\cite{momeni2022fairblock}.

FairBlock, proposed in~\cite{momeni2022fairblock}, keeps this structure but decrypts a whole block with one key. It uses distributed identity-based encryption (IBE), in which a master key derives a private key for any identity, and it takes the block height as that identity. A committee of keepers holds the master key jointly through a one-time distributed key generation (DKG), so that no single keeper can decrypt. Each transaction is encrypted under the height of the block it targets and carries a hash commitment to its plaintext. Validators therefore fix the order over ciphertexts they cannot read, and fixing the order before any content is revealed is what prevents front-running. Once the block is ordered, more keepers than an adversary can corrupt each contribute a share of that block's key, and a relayer combines the shares into the key and decrypts the entire block at once. Because one key decrypts every transaction of the block, shares are exchanged only once per block, which reduces the message count from $N \times S$ to $S$. The saving is in the number of messages, not in the size of each. In the work's benchmarks, which cover the distributed IBE alone rather than a running chain's throughput or latency, extracting a block key with 100 keepers took 147.39~ms on average over 100 runs on a single core~\cite{momeni2022fairblock}. With 1000 transactions and 1000 keepers, of which two-thirds send shares, the IBE shares total 170.8~KB, against 42.7~MB of partial decryptions under threshold decryption~\cite{momeni2022fairblock}.

In the consensus-level variant of FairBlock, one of its two deployments, the validators themselves act as keepers and relayers~\cite{momeni2022fairblock}. The work prototypes this variant as a chain built on the Cosmos SDK, whose validators submit their key shares as ordinary transactions. It also proposes a more efficient form, in which validators carry their shares as ABCI++ vote extensions and the next proposer combines them, decrypts the block, and includes the plaintext transactions in the next block~\cite{momeni2022fairblock}.

The other deployment runs FairBlock as smart contracts that any decentralized application can adopt, and the work prototypes this contract deployment in Solidity~\cite{momeni2022fairblock}. Its keepers and relayers are any parties that enroll with a deposit, such as users, validators, or oracle networks, instead of the validator set. A \emph{commitment contract} stores the encrypted transactions and their commitments in the order received, so this contract preserves the order in which the chain delivered them. Once relayers have decrypted these transactions, a \emph{processing contract} checks each plaintext against its commitment and passes the transactions in that order to the \emph{target contract}, the contract of the decentralized application that adopted FairBlock~\cite{momeni2022fairblock}. In both deployments, the work renegotiates the mempool design choice of the baseline, by having the proposer order transactions whose contents are hidden, in favor of ordering fairness.

The work assumes that a majority of the keepers are correct and that the adversary is computationally bounded, adaptive, and Byzantine~\cite{momeni2022fairblock}. In the consensus-level variant, where validators receive key shares in proportion to their stake, the work instead allows the adversary at most $t \le (n-1)/3$ of the $n$ keepers' shares, in line with the two-thirds assumption of the underlying consensus~\cite{momeni2022fairblock}. Decryption also needs at least one correct relayer~\cite{momeni2022fairblock}. The liveness of the system as a whole additionally rests on the underlying consensus, and hence on partial synchrony in a Tendermint chain.

Under this model, the work supports two properties, proving one and arguing for the other. It proves transaction confidentiality, which means that a ciphertext reveals nothing about its transaction before the key of the block it targets is released. The proof is a reduction to the security of the DKG and of the IBE, which holds in the random-oracle model under the Bilinear Diffie-Hellman assumption~\cite{momeni2022fairblock}. For order-fairness, its term for ordering fairness, the work instead adopts the formal definition of prior work and argues from the order of execution. The argument is that transactions execute in the order in which their commitments were recorded. The commitment contract records the commitments in the contract deployment, and the chain itself records them in the consensus-level variant. In both deployments, no party in the decryption and execution phase, the validators included, can change the recorded order. For the contract deployment, the argument adds that no party can insert a transaction ahead of the decrypted batch, because the target contract accepts only transactions delivered by the processing contract~\cite{momeni2022fairblock}.\footnote{In our reading, this argument shows that the recorded order cannot change, not that it matches the order required by the definition the work adopts, under which a transaction that a large fraction of the nodes received before another is not executed after it~\cite{momeni2022fairblock}. The proposer that orders the commitments still chooses that order, although it cannot read the transactions.}

\subsubsection{Making long-range attacks slashable by importing Bitcoin's proof of work}
\label{subsec:babylon}

Slashing punishes a validator only through its bonded stake (Section~\ref{subsec:choice-bonding}). Once a validator has unbonded and withdrawn its stake, its signing keys are no longer backed by stake, so nothing remains to slash. An adversary who later acquires such keys, once held by validators with more than two-thirds of the voting power at a point deep in the past, can sign an alternative history from that point. Because a chain's block timestamps are set by its quorum, the adversary can stamp this forged history with past times, so its timestamps look as legitimate as the honest chain's. A node that joins late then sees two validly signed chains, with no way from the chain data alone to tell which is canonical. This is the long-range, or posterior-corruption, safety attack (Section~\ref{subsec:conditional-measured})~\cite{tas2022babylon}.

The construction proposed in~\cite{tas2022babylon} first shows that this attack is not an artifact of any particular design. It proves that no proof-of-stake protocol can achieve slashable safety without an external source of trust. Such a source supplies what the validators cannot generate on their own, namely an unforgeable record of which history was published first. A latecomer consults that record, adopts the history that appeared earliest, and discards the alternative published later. Because this record cannot come from within the protocol, the work imports the missing trust from Bitcoin.

That trust is imported through Babylon, the work's name for a proof-of-work chain of its own, distinct from Bitcoin~\cite{tas2022babylon}. Bitcoin's miners merge-mine it, producing its blocks with the same work they already spend on Bitcoin, so it draws on Bitcoin's hash power at no additional energy cost.\footnote{In merge mining, a Bitcoin miner hashes blocks that contain both Bitcoin and Babylon transactions, so that each hash it computes serves both chains. A hash that falls in Babylon's own target range yields a Babylon block, under a criterion different from the one for a Bitcoin block~\cite{tas2022babylon}.} Babylon offers proof-of-stake chains a timestamping and checkpointing service. Proof-of-stake nodes post checkpoints to Babylon, each a hash that commits to the consecutive blocks finalized since the previous checkpoint. A single correct node posting checkpoints suffices. Babylon's miners timestamp each checkpoint and accept it only after checking that the blocks behind it are available, by retrieving or sampling them, without interpreting any proof-of-stake semantics. This check is what gives the timestamp meaning. It forces a checkpoint to commit to data that is genuinely public, so that an attacker cannot timestamp a hidden history early and reveal it only later, once its checkpoint already appears to predate the honest chain's. A plain timestamp on an existing chain cannot perform this availability check, so the design uses a dedicated proof-of-work chain whose miners do, rather than posting checkpoints directly on Bitcoin or Ethereum~\cite{tas2022babylon}. Babylon's timestamps drive a fork-choice rule. When the proof-of-stake chain forks, the canonical branch is the one whose first checkpoint carries the earlier Babylon timestamp. Stake withdrawal is likewise tied to Babylon, in that a validator may withdraw its stake only after the checkpoint covering the block that contains its withdrawal request is buried to a chosen depth in Babylon and no fraud proof, evidence that the validator double-signed, has appeared against it; otherwise it is slashed.

These rules give the protection against long-range attacks that the baseline does not provide, and they rely on Bitcoin's proof of work to do so. An adversary who mounts a safety attack and wishes to avoid slashing must instead reorganize the Babylon blocks beneath the honest checkpoint. Because Babylon is merge-mined by Bitcoin's miners, reorganizing those blocks takes a majority of Babylon's hash power~\cite{tas2022babylon}. Safety violations by equivocation, which the baseline's accountable safety covers (Section~\ref{subsec:accountability}), thus stay slashable even in a long-range attack, as long as no adversary controls a majority of Babylon's hash power, which is tied to the hash power of Bitcoin's miners.

Babylon also makes two kinds of misbehavior attributable, and hence slashable, that the baseline does not attribute, namely transaction censorship (Section~\ref{subsec:conditional-measured}) and stalls (Section~\ref{subsec:liveness}), which the work treats as failures of chain quality and of chain growth~\cite{tas2022babylon}. In each case, nodes post more than checkpoints to Babylon. For censorship, a node submits the censored transaction itself, whose Babylon timestamp and availability check make it provably public and dated, so no validator can later claim it never existed or could not see it. Validators that propose or vote for a block that omits it are slashed, once that block is checkpointed on Babylon after a grace period that follows the submission~\cite{tas2022babylon}. For a stall, a node posts evidence of the stall, and the validators then publish the next round's proposals and votes on Babylon rather than over the network. A validator whose proposal or vote does not appear there within the grace period is slashed~\cite{tas2022babylon}. Because the construction ties stake withdrawal to Babylon and extends slashing to censorship and stalls, it renegotiates bonding and slashing and moves economic security, accountability, and censorship resistance.

The work proves that these guarantees hold under a synchronous network, in which every message among correct validators, Babylon's miners and the other nodes of the proof-of-stake chain arrives within a known delay bound. They also require at least one correct node of the proof-of-stake chain at all times, which need not be a validator, but they make no honest-majority assumption on the proof-of-stake validators. Table~\ref{tab:matrix} marks Babylon's cells $\oplus$~[sync, Bitcoin], because the guarantees hold only under a synchronous network and the security of Bitcoin's proof of work.

The work notes that the grace period sets the worst-case finalization latency while censorship or a stall is under way. A longer grace period, however, makes the attack costlier to mount without being slashed. For new chains whose stake is still worth little, and hence cheap to attack, the work also describes an optional bootstrapping mode, in which a block becomes final only once its checkpoint is buried deep in Babylon. This mode gives up the chain's fast finality in exchange for a higher attack cost, and the chain can leave it as its stake grows~\cite{tas2022babylon}.

\subsubsection{Raising liveness resilience to one-half under synchrony by checkpointing onto unmodified Bitcoin}
\label{subsec:bitcoin-pos}

Tendermint keeps liveness only while fewer than one-third of the active validators are adversarial, so its liveness resilience is below one-third. An adversary holding one-third or more can stall the chain~\cite{tas2023bitcoin}. Under a synchronous network, the construction proposed in~\cite{tas2023bitcoin} raises this resilience to below one-half, while keeping accountable safety, by checkpointing the chain onto unmodified Bitcoin. The same checkpoints also make long-range attacks (Section~\ref{subsec:conditional-measured}) slashable and shorten the delay before a validator can withdraw its stake. The work is the second paper in the Babylon line, and it shares its slashable-safety foundation with the earlier Babylon design (Section~\ref{subsec:babylon})~\cite{tas2022babylon}. Three contributions are distinct to this later work.

The first contribution is architectural. The earlier design builds a separate proof-of-work chain, merge-mined by Bitcoin's miners, whose miners check that a checkpoint's block data is available~\cite{tas2022babylon}. This work instead posts checkpoints directly onto Bitcoin as it is, through the \texttt{OP\_RETURN} opcode, and changes nothing about Bitcoin~\cite{tas2023bitcoin}. The chain groups its blocks into epochs. At each epoch's end, its validators sign the epoch's last finalized block, which commits the whole epoch through the hash chain, and a correct validator posts a single checkpoint of it to Bitcoin. The checkpoint is that block's hash alone, the epoch number, and the signatures of more than two-thirds of the active validators. On Tendermint, this quorum is a subset of the pre-commit signatures that already finalized the block~\cite{tas2023bitcoin}. The checkpoint therefore adds no new signing round. It repackages the pre-commit quorum Tendermint already produces.

Posting a checkpoint does not delay the commit. Validators commit each block as the baseline does, without waiting for Bitcoin, and the checkpoints matter only to clients, which use them afterwards to choose the canonical chain. A client is any full node that downloads the blocks and the Bitcoin chain to obtain the canonical chain, whether a validator or a node that joins late~\cite{tas2023bitcoin}. It takes the order of the checkpoints from Bitcoin, prefers the history with the earlier checkpoint, and appends the newest finalized blocks to that checkpointed prefix, so a transaction reaches its chain at Tendermint's latency~\cite{tas2023bitcoin}. Because unmodified Bitcoin cannot check availability, that check also falls to the client. A client that finds a validly signed checkpoint whose block data it cannot retrieve stops extending its chain at that checkpoint, rather than risk a safety violation once the hidden blocks are revealed~\cite{tas2023bitcoin}. Only a supermajority could sign such a checkpoint, so only an adversary holding more than two-thirds of the active validators can freeze the clients' chains in this way.

Bitcoin's order of the checkpoints is also what makes a long-range attack slashable. A history that validators sign after they have withdrawn their stake conflicts with blocks that are already checkpointed, so the clients discard it in favor of the earlier checkpoint. A safety violation can therefore succeed only among recent blocks whose checkpoints are not yet deep in Bitcoin~\cite{tas2023bitcoin}. Withdrawal is held back for that window. A validator's stake stays bonded, and hence slashable, until the checkpoint covering the block that contains its withdrawal request is $k$ blocks deep in Bitcoin and no evidence that the validator double-signed, whether a fraud proof or a conflicting checkpoint carrying its signature, has appeared there~\cite{tas2022babylon,tas2023bitcoin}. The validators behind a safety violation therefore still hold bonded stake when the violation comes to light, whereas without this delay they could withdraw first and sign the conflicting blocks at no cost. Because the construction ties stake withdrawal to Bitcoin, it renegotiates bonding and slashing and moves economic security, as the earlier design does~\cite{tas2022babylon}.

The second and third contributions concern liveness. The earlier Babylon design accepts a stall and penalizes it~\cite{tas2022babylon}. This work instead raises the resilience threshold from below one-third to below one-half of the active validators, which is its second contribution. Its third contribution is a matching impossibility result. Under synchrony, a protocol that provides accountable safety cannot stay live against one-half or more adversarial validators if the protocol writes less than about half of its validators' input data to a data-limited timestamping service such as Bitcoin~\cite{tas2023bitcoin}. A timestamping service is data-limited when it limits the total size of the messages that it records~\cite{tas2023bitcoin}. The checkpoints play a different role on each side of one-third. Below one-third adversarial active validators, the chain runs as in the baseline and commits at its fast proof-of-stake latency, with its epoch checkpoints still posted to Bitcoin in the background. Between one-third and one-half, that fast path can stall, leaving a transaction uncommitted past the proof-of-stake latency. A correct validator that holds such a censored transaction then posts a liveness transaction to Bitcoin, naming the censored transaction by its hash. If the censored transaction is still missing once the liveness transaction is deep in Bitcoin, the validators stop running Tendermint and fall back to rollup mode, in which they bundle transactions for Bitcoin to order directly~\cite{tas2023bitcoin}. In this mode, Bitcoin's order of the bundle checkpoints is the chain's order. Because that ordering is Bitcoin's, not the validators', a bundle needs only a majority to certify, not the two-thirds an in-protocol quorum requires. The correct validators, being a majority, always reach that threshold, while an adversary below one-half can neither forge a valid bundle checkpoint nor freeze the clients' chains with unavailable data~\cite{tas2023bitcoin}. Because the fallback hands the ordering to Bitcoin once Tendermint's rounds stop deciding, the construction also renegotiates locking and round progression and moves liveness.

The work proves these results under a synchronous network, in which every message from a correct validator reaches every other correct validator and every client within a known delay bound, a stronger assumption than the baseline's partial synchrony~\cite{tas2023bitcoin}. For the design without the rollup fallback, the work extends its slashable-safety result and its liveness result below one-third to a restricted form of partial synchrony, in which the messages between validators and Bitcoin's miners always arrive within the delay bound, the messages among validators and clients may be delayed arbitrarily until GST, and clients stay online~\cite{tas2023bitcoin}. The results further require at least one correct active validator at all times, fewer than one-half adversarial active validators for the one-half liveness result, and Bitcoin secure at depth $k$ against an adversary mining below one-half of the hash rate. The thresholds on validators are counts, not shares of voting power, so they carry over to voting power only when every validator holds the same voting power. Table~\ref{tab:matrix} marks both of the work's cells $\oplus$~[sync, Bitcoin], because the results hold only under a synchronous network, at least toward Bitcoin's miners, and the security of Bitcoin's proof of work.

Beyond these proofs, the work measures one use of the design, a shorter stake-withdrawal delay. A proof-of-stake chain holds a withdrawing validator's stake bonded for a fixed, weeks-long unbonding period, long enough to detect and slash misbehavior before the stake leaves. The construction instead releases the stake once the withdrawal's checkpoint is deep enough in Bitcoin, so the delay becomes Bitcoin's confirmation latency rather than a conservative fixed window. To estimate it, a prototype checkpointer posted hourly checkpoints to the Bitcoin mainnet for one day, for an emulated Tendermint chain of 100 validators that signs its votes with BLS, so that each checkpoint's quorum of 67 signatures aggregates into one. At the two fee levels tested, a checkpoint took on average 192.7 and 204.1 minutes to lie 20 blocks deep in Bitcoin. From these confirmation times, the work concludes that the withdrawal delay falls from weeks to below four hours, at an annual cost under \$10{,}000 in Bitcoin fees for posting the checkpoints~\cite{tas2023bitcoin}. This low cost relies on the aggregate signatures, without which it would be on the order of millions of US dollars~\cite{tas2023bitcoin}.

The liveness gain has a latency cost, which the work states. In the fallback, a transaction becomes final only at Bitcoin's confirmation latency, so while the adversary holds between one-third and one-half of the active validators, the chain gives up the baseline's fast finality~\cite{tas2023bitcoin}.

\subsubsection{Extending accountability from safety faults to liveness faults under bounded asynchrony}
\label{subsec:accountable-liveness}

Section~\ref{subsec:accountability} shows that the baseline attributes a same-round equivocation fork. After such a fork, the more than one-third of the voting power that provably double-signed can be slashed, and no correct validator is falsely accused. The work~\cite{lewis2025accountable} asks whether liveness can be made accountable in the same sense. It defines a liveness violation, or stall, as a transaction left uncommitted past a confirmation deadline. \emph{Accountable liveness} then requires that every such violation yield, with high probability, a certificate of guilt naming a guaranteed large subset of the responsible validators, so that their stake too can be slashed, again without accusing a correct validator. The difficulty is that a missing vote does not identify its cause, since a vote that a validator withheld is indistinguishable from one the network dropped. Making this absence attributable is the problem the work addresses.

The construction proposed in~\cite{lewis2025accountable} does not guarantee progress once the adversary controls one-third or more of the validators, but rather makes a lack of progress attributable and slashable. This attribution matters most when the adversary controls at least one-third but fewer than one-half of the validators, because the baseline can then lose liveness, whereas the construction still makes a stall attributable~\cite{lewis2025accountable}. From one-half on, no protocol can attribute a stall, even under synchrony, if it is safe under partial synchrony at least when every validator is correct~\cite{lewis2025accountable}, as the baseline is. The work counts validators rather than voting power, so its bounds match bounds on voting power only when all validators hold equal voting power. The construction changes the baseline in three ways, one in the network model and two in the protocol. First, for accountable liveness, the construction assumes a network model that the work calls $x$-partial-synchrony, which is stronger than the baseline's partial synchrony~\cite{lewis2025accountable}. Time is divided into periods of a fixed length. A period, or a Tendermint round, which the work calls a view, is synchronous when every message sent during it arrives within a network-delay bound $\Delta$, and asynchronous otherwise. Before GST, every long enough interval has at most a fraction $x$ of asynchronous periods. Partial synchrony is the case $x = 1$. Second, each round gains a third vote, the \emph{VoteLive} vote. A validator casts it late in each round if the transactions it held at that round's start have committed by then. These votes serve only to detect and attribute stalls, and no validator waits for them to propose, prevote, precommit or commit~\cite{lewis2025accountable}. Third, the delays between a round's steps grow from one $\Delta$ in the work's clock-driven version of Tendermint to two or three.\footnote{In that version, every step of a round starts at a fixed time, one $\Delta$ after the previous step~\cite{lewis2025accountable}. In the baseline, a validator instead moves to its next step once it holds the votes that the step needs or once that step's timeout expires (Section~\ref{subsec:core-mechanics}).} Liveness alone does not need this lengthening, but it lets a correct validator tell from the messages it has received which messages every other validator had received in time, and hence which votes each validator owed~\cite{lewis2025accountable}.\footnote{Under synchrony, a correct validator re-broadcasts every message it receives. A message that it has received by time $t$ therefore reaches every validator by $t + \Delta$, and a message that it has not received by $t$ had reached no correct validator by $t - \Delta$. Most steps wait $2\Delta$, one $\Delta$ for each direction, and the two steps that wait $3\Delta$ add one more, because a quorum of prevotes, and later a quorum of precommits, needs one more hop to reach every validator. With these bounds, a validator can tell both which vote another validator owed and whether that validator could have had an excuse for withholding it, such as a lock on another block, so a missing vote is blamed only when its inputs must have arrived and no excuse was possible~\cite{lewis2025accountable}.}

The VoteLive votes make a stall detectable. In the normal case, more than two-thirds of the validators cast a VoteLive vote in every round. Detection examines a \emph{window}, which is a fixed number of the most recent rounds, with its length set in advance from the parameters of the $x$-partial-synchrony model. A stall is signaled when the VoteLive votes fall short of a more-than-two-thirds quorum in every round of this window.

A signaled stall is attributed from the consensus and VoteLive votes over that window. The evidence for this attribution comes from \emph{records} that each validator signs and broadcasts on a continuing basis. A validator's record holds the blocks and votes that it has received from other validators, each marked with the time of its arrival. The attribution rests on a synchronous round with a correct proposer. The $x$-partial-synchrony model ensures that most of the window's rounds are synchronous. The work's analysis assumes that each round's proposer is chosen uniformly at random, without the baseline's weighting by voting power, so many of those rounds also have a correct proposer. In such a round, messages arrive on time, so every transaction pending at a correct validator early in the round has reached all validators, and the correct proposer's valid block includes it. If that block still fails to commit, the records show some of its prevotes or precommits missing. Since no message is lost in such a round, each missing vote was necessarily withheld, unless its validator was legitimately locked on an earlier block. On this basis, a validator that detects the stall applies an adjudication rule to the records. A record blames a validator for a round when it shows a vote that this validator owed in that round but did not cast. The rule divides the window into groups of consecutive rounds and outputs each validator that most records blame in more than a set fraction of these groups. The detecting validator broadcasts an accusation against each validator that the rule outputs. Accusations against one validator from more than one-half of the validators form a certificate of guilt against it, and that validator is held responsible for the stall~\cite{lewis2025accountable}. No correct validator is held responsible. In a synchronous round, a correct validator's votes reach everyone, so no correct validator's record blames another correct validator for its prevotes or precommits. Nor does a correct validator's record blame a correct validator for withholding its VoteLive vote while that validator's pending transactions have not committed. A validator counts that vote as missing only when its own record shows that every transaction it held shortly after the round began has committed within the round, and in a synchronous round these include the blamed validator's pending transactions, which have therefore committed in time too. Blame can still arise in asynchronous rounds, but when $x < 1/2$, too few groups of rounds are asynchronous for a correct validator to be blamed in the set fraction of them~\cite{lewis2025accountable}. Because the construction makes a stall attributable, and hence slashable, through a VoteLive vote and longer delays between the steps of each round, it renegotiates locking and round progression as well as bonding and slashing, and it moves accountability but worsens latency.

The work establishes its results by proof. Under $x$-partial-synchrony, accountable liveness is achievable if and only if the adversary controls fewer than one-half of the validators and the network is boundedly asynchronous, $x < 1/2$~\cite{lewis2025accountable}. Because bounded asynchrony is a strictly stronger assumption than the baseline's partial synchrony, Table~\ref{tab:matrix} marks both of the work's accountability cells $\oplus$~[bounded async].

The construction has two costs. The first is latency. The lengthened delays between steps are fixed, i.e.,\ $2\Delta$--$3\Delta$, so a round lasts $12\Delta$~\cite{lewis2025accountable}. The baseline, by contrast, moves to its next step as soon as the proposal arrives or as soon as votes for one value from more than two-thirds of the voting power arrive. It waits out a timeout only when the proposal does not arrive in time, or when votes from more than two-thirds of the voting power arrive but are split across values~\cite{buchman2018latest}. Latency therefore worsens, because the delay bound $\Delta$, not the actual network delay, paces each round. The second is the communication overhead that the broadcast records add, and the work leaves reducing that overhead to future work, because it aims to show that accountable liveness is achievable, not to give its most efficient implementation~\cite{lewis2025accountable}.

Two further directions remain. First, the construction can be implemented and evaluated empirically on a running engine. Second, it assumes that a correct proposer's block carries all of a correct validator's pending transactions. Under a bounded block size, a backlog of pending transactions larger than one block cannot commit within a round, so correct validators do not cast their VoteLive votes. If the backlog persists through every round of the window, the construction signals a stall while the chain is still making progress. The work already gives each transaction its own confirmation deadline, a fixed time after the transaction's input, but the VoteLive condition does not consult these deadlines. A validator could instead cast its VoteLive vote if every pending transaction whose deadline has passed has committed by then, rather than only if its whole pending set has committed. This condition would remove the false stall signal, because a correct validator would then miss its vote only when a transaction is past its deadline. That case is what the work's definition of a liveness violation counts.

\subsubsection{Fair reward distribution through a growing commit timeout}
\label{subsec:correctness-fairness}

Section~\ref{subsec:conditional-measured} shows that the baseline is fair in reward distribution only if the commit timeout is long enough for late precommits to arrive, the next proposer records them all, and the reward rule pays every recorded participant. Against a slow validator's precommit missing the record, the baseline relies on the commit timeout, a fixed wait before the next height (Section~\ref{subsec:reward-distribution}). The point of fairness is that a correct process holding a tenth of the stake earns about a tenth of the total reward~\cite{amoussou2018correctnessext}, which fails whenever a slow correct validator goes unpaid or, in a case the work adds to Section~\ref{subsec:conditional-measured}'s threats, a validator that took no part in deciding the block is paid anyway. The work~\cite{amoussou2018correctnessext} asks whether any timeout, or any reward mechanism at all, can rule out such failures.\footnote{Section~\ref{sec:background} already draws on the work's correctness analysis of the original algorithm for its safety and liveness arguments (Sections~\ref{subsec:safety} and~\ref{subsec:liveness})~\cite{amoussou2018correctness}. The analysis of rewards is the work's contribution beyond the baseline.}

To answer it, the work formalizes the baseline's reward mechanism and finds it unfair in two cases~\cite{amoussou2018correctnessext}. Its model is the baseline's partial synchrony, which it calls eventual synchrony, with fewer than one-third of the $n$ validators, $f < n/3$, Byzantine. Its bounds are stated in validators rather than voting power, so they apply to voting power only when all validators hold equal voting power. Its notion of fairness extends fairness in reward distribution (Section~\ref{subsec:conditional-measured}) by one requirement, namely that no faulty validator or other non-participant may be rewarded. It also calls a reward mechanism \emph{eventually fair} when fairness holds from some height on. Its formalization of the mechanism differs from the participation record of Section~\ref{subsec:reward-distribution}. Once a block is decided, each validator broadcasts a \emph{commit message},\footnote{The work models the commit as a separate message type, which its reward mechanism keys on. CometBFT, the current implementation, has no separate signed commit message. The precommits that committed a block instead form the next block's participation record, its \statevar{LastCommit} field (Section~\ref{subsec:reward-distribution})~\cite{cometbft}.} an announcement of the decision that is distinct from its precommit vote. Each validator collects these messages until a commit timeout expires, and the proposer of the next height rewards the validators whose commit messages it has collected. In the first case, a slow but correct validator whose commit message arrives after the timeout goes unrewarded. The work proves that the mechanism is not eventually fair when the timeout is shorter than the maximum message delay, because such a validator can then go unrewarded at any height, including after the network becomes synchronous~\cite{amoussou2018correctnessext}. The second case, which the work identifies, involves a free-rider, a validator that sent no proposal or vote while the block was being decided~\cite{amoussou2018correctnessext}. Because a commit message only announces the decision, such a validator can observe that the block was decided, broadcast a commit message like any other validator, and thereby free-ride into the reward set.

The work proposes a two-part fix, one for each case. A modulable commit timeout that grows until it catches up with the message delay fixes the late-commit case. For the free-rider, each commit message is extended so that its sender records which validators it observed taking part in each phase, and, as one example, the rewarding proposer credits a validator only when at least $f+1$ commit messages name it as a participant. Because at most $f$ validators are faulty, those $f+1$ attestations include at least one from a correct validator, so a validator that took no part is never vouched for and is excluded. When commit messages suffice to detect a validator that did not send its expected messages, the work proves that the modified mechanism is eventually fair~\cite{amoussou2018correctnessext}. If, in addition, validators are selected in proportion to stake, which the work calls a fair selection mechanism, the work's repeated-consensus protocol with the modified mechanism is eventually fair as a whole~\cite{amoussou2018correctnessext}. Because the modified mechanism changes both how long validators collect commit messages and which validators may enter the reward set, it renegotiates reward distribution and moves fairness in reward distribution.

The reward-fairness result holds under two conditions. The first is that fairness holds only eventually, which the work shows is the best possible under the baseline's partial synchrony. For any blockchain built on repeated consensus, not only for Tendermint's mechanism, a fair reward mechanism exists if and only if the system is synchronous, and an eventually fair one exists if and only if the system is eventually synchronous or synchronous~\cite{amoussou2018correctnessext}. The impossibility rests on the fact that fairness forces the rewarding validators to know exactly which validators were correct. That knowledge requires them to have received all of those validators' messages before granting the reward, which a system that is not synchronous does not guarantee. Under partial synchrony the delay bound holds only after GST, so the growing commit timeout catches up with the message delay only eventually, and the mechanism attains eventual rather than full fairness. The second condition concerns the deciding proposer. The work's analysis does not treat a Byzantine deciding proposer, and its proof of the eventual result considers only a correct one. The reward set is that proposer's own local set, which enters the consensus instance together with its block. The work does not require the block-validity predicate, which is application-dependent, to check this set, so a Byzantine proposer whose block is decided can fill it arbitrarily. This case can recur. The work's round-robin proposer function gives every validator its turns, and its model bounds only how many validators are Byzantine at each height, not how often they propose. The model therefore admits executions in which Byzantine proposers decide blocks at infinitely many heights. The work's theorem that its protocol with the modified mechanism is eventually fair assumes that, from the commit messages, validators can distinguish correct from Byzantine behavior~\cite{amoussou2018correctnessext}. In our reading, this hypothesis does not by itself require correct validators to reject a block whose reward set is wrong, and unless they reject such a block, fairness can fail at each height where a Byzantine proposer's block is decided, so eventual fairness can fail too.

\subsection{Synthesis of the surveyed works}
\label{subsec:map-shows}

Reading the classification of Table~\ref{tab:matrix} along its property columns, we find that the
surveyed works fall into groups by their aim. The first group improves the performance of the
baseline, whether its throughput, its latency or its scalability. The second group extends
accountability to actions that harm the network, whether by depriving other validators of their
rewards, by censoring transactions or votes, or by stalling the chain. The third group protects the information that the baseline exposes or
makes publicly available, so that validators and outside observers cannot turn that information to
their advantage.

The works that improve the performance of the baseline focus on the sources of delay that lie on the
critical path of block production. The sources they address are \emph{gossip amplification} and
the sequential decision of heights. We define gossip amplification as the message overhead of
disseminating votes by gossip. This overhead comes on top of the quadratic number of vote deliveries
that each round needs (Section~\ref{subsec:complexity}) and arises from duplicate deliveries
in push gossip, in which a validator can receive the same vote from different
peers~\cite{cason2025boosting} (Section~\ref{subsec:choice-gossip}). The time that a validator spends processing gossiped votes, duplicates included, until it holds a quorum adds
to the duration of each consensus step and so to the time to produce each block. Filtering
the votes that no longer contribute to progress once a quorum is observed and bundling the votes
that are still relayed into fewer messages, both of which make gossip consensus-aware, are methods
pursued by related work to reduce gossip amplification~\cite{cason2025boosting}. The second source, the
sequential decision of heights, makes each block wait for the voting on its
predecessor~\cite{buchman2018latest}. The time that validators spend waiting for each step's votes could already
serve the next block~\cite{lei2020chained}.
Pipelining heights puts this time to use by overlapping
the voting on one block with the proposal and execution of the next. It improves throughput but
defers each block's commit by several heights~\cite{lei2020chained}.

The works that extend accountability aim to make validators accountable for as many of their actions
as possible. The baseline provides accountability for equivocation, a safety deviation whose
evidence is the conflicting signatures (Section~\ref{subsec:accountability}). For liveness, by
contrast, it provides none. A liveness violation, or stall, can arise from omissions such as
withheld votes, and an omission leaves no evidence on which to slash the validators that cause the
stall. Accountability for either guarantee matters only beyond the baseline's fault bound, within
which both guarantees hold (Section~\ref{sec:background}). Even within the bound, however, the
baseline leaves some actions to the validators' discretion, and a validator can use them to harm the
network without violating safety or liveness. A proposer may, for instance, drop a correct
validator's precommit from the participation record, which counts toward that validator's
downtime~\cite{cosmossdk} and, under a reward rule that pays only included precommits, costs it its
reward (Section~\ref{subsec:reward-distribution}). The application's downtime
penalty (Section~\ref{subsec:reward-distribution}) applies when a validator misses more than a set
share of votes over a window of blocks, rather than on evidence~\cite{cosmossdk}. Under partial
synchrony, the exclusion of a precommit is, like a
withheld vote, indistinguishable from message delay~\cite{haeberlen2007peerreview}. A rule that
holds a proposer accountable for such an exclusion should nonetheless minimize, and ideally
eliminate, the chance of penalizing a correct proposer that did not receive the vote. In response, the works of this group augment the
baseline so that actions that can harm the network, whether deviations or discretionary choices,
fall under enforceable rules backed by verifiable evidence, are overridden, or are prevented.

Identifying a withheld vote or an exclusion depends on the local view or state of a validator, that
is, on what it sent and what it received. A first path to attribution makes that view or state
publicly verifiable by publishing records of it. For votes, such a published record has been the receiver's
signed acknowledgment of receipt, given as a reply to each vote and reported to designated
validators~\cite{xian2019improved} or as a
running transcript of the messages that the receiver has seen~\cite{lewis2025accountable}.
Publishing such records adds messages~\cite{xian2019improved,lewis2025accountable}, but it supplies
the evidence that attribution requires. The published records also make validators'
views of the network comparable, and comparing those views enables attributing misbehavior such as
censorship~\cite{xian2019improved} or a stall~\cite{lewis2025accountable}.

Instead of records exchanged among the validators, a second path to
attribution posts the omitted information to a witness chain, an outside chain that all validators
can read. An omission then becomes attributable, because the witness chain dates the data that nodes post to
it and so fixes an order of events that the validators cannot rewrite~\cite{tas2022babylon}. A
valid transaction that proposers repeatedly omit thus becomes
provably public when a node posts it together with a censorship complaint. Once
the complaint is deep enough in the witness chain for every correct validator to have observed it,
any validator that proposes or votes for a block that omits the transaction becomes
slashable~\cite{tas2022babylon}. Beyond slashing, later
work in the same line proposes handing off the ordering to the witness chain to override the
censorship if it persists~\cite{tas2023bitcoin}. The order of
events that the witness chain fixes also improves economic security against the long-range attack,
because at a fork the canonical chain is the one whose blocks were ordered and timestamped
first~\cite{tas2022babylon,tas2023bitcoin}.

Some works have identified that a certain level of accountability cannot be achieved without a
timing model stronger than the baseline's partial synchrony. Under partial synchrony, accountable
liveness is proved impossible for an optimally resilient protocol such as the
baseline~\cite{lewis2025accountable}, and fairness in reward distribution is proved attainable only
eventually~\cite{amoussou2018correctnessext}. Such fairness rests on the same knowledge as
accountable liveness, namely whose votes are missing and whether each was withheld or delayed.
Telling the two apart requires a bound on message delay, which must hold at all times for full
fairness~\cite{amoussou2018correctnessext} and in most rounds for accountable
liveness~\cite{lewis2025accountable}. Going beyond partial
synchrony is therefore inevitable for accountable liveness and for full rather than eventual
fairness, and doing so opens a third path to attribution. This path assumes such a model for
attribution, either synchrony~\cite{xian2019improved,tas2022babylon} or bounded
asynchrony~\cite{lewis2025accountable}, and it underlies the other two paths, since each of these
works combines it with published records or a witness chain. Such a model assumes a known bound on message delay that holds at all times, or
at least most of the time, so a message still missing past that bound can be attributed to its
sender rather than to the network.

Going one step further, some designs prevent deviations rather than make them attributable. Such
work, focused on equivocation, deploys trusted hardware that never attests two conflicting messages
for the same consensus step, which raises the tolerated share of Byzantine validators from below
one-third to below one-half~\cite{beltrando2023tendertee}.

The works that protect information propose countermeasures to hide the information that is publicly
available in the baseline and can be exploited to an actor's advantage. The baseline makes two kinds of such
information public early enough to exploit, and each exposure threatens a property. The first kind
is the proposer schedule. Because the schedule is predictable, anyone can target an upcoming
proposer, for instance with a denial-of-service attack, and harm proposer
availability~\cite{buchman2022revisiting}. Seeding proposer selection with the output of a
verifiable random function mitigates such targeting~\cite{bui2023enhanced}.\footnote{As
Section~\ref{subsec:conditional-measured} describes, deployments also hide each validator behind
sentry nodes, without changing the protocol~\cite{buchman2022revisiting}.} The second kind is the
content of pending transactions. Because pending transactions are readable to every node that relays them, an observer
can front-run a transaction before its order is fixed and harm ordering
fairness~\cite{momeni2022fairblock}. As long as fewer key holders than the decryption threshold collude, encrypting transactions
counters such front-running, because
the content of a transaction that the block includes is revealed only after its order is fixed, too late for a front-runner to
act on it~\cite{momeni2022fairblock}. Both countermeasures thus reveal the information later
than the baseline does, yet they differ in how fully they protect. Unlike an encrypted transaction,
a seeded schedule reveals the proposer of every round of a height by the time the previous block
commits, so an attacker can target the proposers of later rounds well before their turns come
(Section~\ref{subsec:vrf-proposer}). Hiding therefore protects more fully when the reveal comes
after the event that an attacker could disrupt with the information.

\section{Conclusion and Future Directions}
\label{sec:conclusion}
\label{sec:open}

The Tendermint protocol, through its engines, underpins many application-specific
proof-of-stake chains, yet broader surveys mention it only alongside many other protocols. We therefore
devoted this survey to Tendermint and the works that propose changes to it, as one family. We
first identified the design choices through which Tendermint's baseline delivers its guarantees
and its other properties, along the vote path, the proposal path, and the economic layer
(Section~\ref{subsec:dna-choices}). Along the vote path, the choices govern how a vote is
disseminated, verified, and counted toward a lock or a commit, and through them the baseline
provides its safety, liveness, and accountability for equivocation. Along the proposal path, they
determine which validator builds each block and which transactions that block contains, and in the
economic layer they govern economic security and fairness in reward distribution. We
then classified the surveyed works by the choice each renegotiates, the properties it moves
(strengthens or weakens), and the fault and timing model its result requires
(Section~\ref{sec:matrix}). Finally, looking across that classification as a whole, we found that
the solutions proposed in the surveyed works cluster around a small number of threads of concern
about the baseline. Within each thread, their approaches draw on a shared set of recurring
strategies (Section~\ref{subsec:map-shows}).

From the classification and from the baseline itself, we draw directions for future work. Unless a
direction states otherwise, we pose it under the baseline's own model of partial synchrony and the
fault bound, which limits Byzantine validators to less than one-third of the voting power.

\paragraph{Attributing a proposer's omission of a precommit}
A proposer in the baseline can leave a correct validator's precommit out of the participation record
(Section~\ref{subsec:reward-distribution}), which harms censorship resistance for votes and, under a
reward rule that pays only included precommits, fairness in reward distribution. Attributing such an omission needs evidence that the proposer
received the precommit and a public record of that evidence that no one can dispute or backdate. Two
surveyed works, an audit of reported vote sets~\cite{xian2019improved}
(Section~\ref{subsec:vote-censorship}) and the Babylon design of~\cite{tas2022babylon} (Section~\ref{subsec:babylon}),
each supply one of these pieces. The audit supplies the evidence, using each receiver's signed vote
reply against a proposer that acknowledged votes and left them out. Its judgment, however, needs synchrony and a coalition that leaves out many validators' precommits, and it relies on correct auxiliary validators, since the work does not consider a faulty one. Babylon supplies the record, since its witness chain, an outside chain that all
validators can read, dates posted data with no honest-majority assumption on the validators. It
posts votes there, however, only to attribute stalls. Yet a gap remains, since neither work alone
attributes a single proposer's omission of one validator's precommit, and no surveyed work, under the
synchrony that both assume, posts the signed acknowledgement to the witness chain. Beyond this combination, a broader open problem is whether a reward mechanism can be specified and proved eventually fair when Byzantine proposers decide blocks. In our reading, the eventual-fairness result of~\cite{amoussou2018correctnessext} does not treat that case (Section~\ref{subsec:correctness-fairness}).

\paragraph{Attributing stalls under a bounded block size}
The work in~\cite{lewis2025accountable} (Section~\ref{subsec:accountable-liveness}) proposes a method to
attribute stalls, such as transactions censored past their confirmation deadline. Its construction,
however, assumes that a correct proposer's block carries all of a correct validator's pending
transactions. Detection is tied to a VoteLive vote that a correct
validator casts in each round only if every transaction pending at the start of the round has committed
within it, and the absence of these votes signals a stall. Yet with bounded blocks, a backlog that
persists across the window of recent rounds that detection examines withholds these votes in each of them and
signals a stall while the chain progresses. Extending this work to a limited block space, a
setting closer to real deployments, remains open under the work's assumptions of bounded
asynchrony and fewer than one-half of the validators adversarial.

\paragraph{Attributing the amnesia attack}
For forks, the baseline already attributes equivocation, because two conflicting signed votes are
evidence that any party can verify. It does not attribute the amnesia attack, in which a validator
that has locked a block prevotes a different block in a later round without the polka that would
justify releasing the lock (Section~\ref{subsec:accountability})~\cite{buchman2022revisiting}.
Nothing
in the signed votes distinguishes a forgotten lock from a legitimate release. The gap matters past
the fault bound, since a fork requires more than one-third of the voting power to be Byzantine, and
the baseline does not attribute a fork produced by amnesia to the validators responsible for it.
A 2022
retrospective, inspired by work on BFT protocol forensics~\cite{sheng2021forensics}, sketches a
transformation of Tendermint in which every prevote carries some information about the proposal
message that triggered it, so that amnesia would become the already-attributable
equivocation~\cite{buchman2022revisiting}. The sketch does not specify what that
information is or how an unjustified prevote would be exposed as invalid, and the retrospective
defers the formal transformation to future work~\cite{buchman2022revisiting}. Whether such a
transformation can be specified and proved to make every amnesia attack attributable remains open,
as do its cost to the vote message and the fault and timing assumptions it needs.

\paragraph{Improving proposer availability within a height}
The work in~\cite{bui2023enhanced} (Section~\ref{subsec:vrf-proposer}) proposes seeding proposer
selection from a verifiable random function to improve proposer availability, so that proposers can
no longer be predicted beyond the next height. Its gain, however, does not extend inside a height. One seed
fixes every round proposer of a height, including those that take over after a timeout, so an
attacker can target them before their turns come. The first height's seed is likewise a public
constant. The seed also comes from a block's proposer, which
the work argues can influence it only by abstaining, at the risk of a slash~\cite{bui2023enhanced}.
What remains open is whether a proposer-selection rule can keep the round proposers of a height
unpredictable, the first height included, with a seed that no single proposer can bias, by
abstaining or otherwise. Such a rule should still let validators find each round's proposer without
exchanging messages, as in the baseline.

\bibliographystyle{ACM-Reference-Format}
\bibliography{references}

\end{document}